\documentclass{article}
\pdfoutput=1
\usepackage{amsmath}
\usepackage{amsfonts}
\usepackage{latexsym}
\usepackage{amscd}
\usepackage{amssymb}
\usepackage{graphicx}
\usepackage{ctable}
\usepackage{float}
\usepackage{tabularx}
\usepackage{color}
\usepackage{subfigure}
\usepackage{graphics} %% add this and next lines if pictures should be in esp format
\usepackage{graphicx} 
\usepackage{epsf}
\usepackage{epsfig}
\usepackage{cancel}%cancel permet en mode Mathématiques de barrer un texte

\newcommand{\R}{\mathbb{R}}

\restylefloat{float}

\title{A Model of the 2014 Ebola Epidemic in West Africa\\ 
with Contact Tracing}

\begin{document}

\maketitle
\author{
\centerline{\scshape Cameron Browne, Vanderbilt University, Nashville, TN 37240, USA}
\vspace{.1in}
\centerline{\scshape Xi Huo, Ryerson University, Toronto, Canada}
\vspace{.1in}
\centerline{\scshape Pierre Magal, University of Bordeaux, Bordeaux, France}
\vspace{.1in}
\centerline{\scshape Moussa Seydi, chef du service des Maladies infectieuses de l'hôpital Fann, Dakar, S\'en\'egal}
\vspace{.1in}
\centerline{\scshape Ousmane Seydi, Ecole Polytechnique, Thi\`es, S\'en\'egal}
\vspace{.1in}
\centerline{\scshape Glenn Webb, Vanderbilt University, Nashville, TN 37240, USA}
}
\vspace{.1in}

\begin{abstract}
A differential equations model is developed for the 2014 Ebola epidemics in Sierra Leone, Liberia, and Guinea. The model describes the dynamic interactions of the susceptible and infected populations of these countries.  The model incorporates the principle features of contact tracing, namely, the number of contacts per identified infectious case, the likelihood that a traced contact is either incubating or infectious, and the efficiency of the contact tracing process.The model is first fitted to current cumulative reported case data in each country. The data fitted simulations are then projected forward in time, with varying parameter regimes corresponding to contact tracing efficiencies. These projections quantify the importance of  the identification, isolation, and contact tracing processes for containment of the epidemics.
 \end{abstract}

\vspace{.1in}

\section{Introduction}

Our objective is to develop a mathematical model of the 2014 Ebola epidemic in West Africa. The model consists of a system of ordinary differential equations for the compartments of the epidemic population. The model incorporates the unique  features of the Ebola outbreaks in this region. These features include the rates of transmission to susceptibles from both infectious cases and improperly handled deceased cases, the rates of reporting and isolating  these cases, and the rates of recovery and mortality for these cases. The model also incorporates contact tracing of reported infectious cases,  and analyzes the efficiency of the  identification and isolation of these cases, and the efficiency of contact tracing measures. 

We apply the model to Sierra Leone, Liberia, and Guinea, first fitting WHO data for each country from outbreak  in the spring of 2014 to September 23, 2014. We then simulate forward projections of the epidemic in each of these countries, based on varied efficiencies in identifying, isolating, and contact tracing of infected individuals. Our model predictions indicate that the containment of the epidemic requires a high level of both the general identification and isolation process and the contact tracing process for removing  infected individuals from the susceptible population.  

\vspace{.5cm}

\section{The Model}

The model, which is of SEIR form (\cite{Althaus},\cite{Chowell},\cite{Legrand},\cite{Lekone}), incorporates specific features of contact tracing in the current epidemics. The model consists of the populations at time $t$ of susceptibles $S(t)$ (capable of becoming infected), exposed $E(t)$ (incubating infected),
$I(t)$ (infectious infected), contaminated deceased $C(t)$ (improperly handled corpses of infected), isolated infectious $II(t)$ (exposed and infectious infected who have been identified and isolated from the susceptible population), and removed $R(t)$ (infectious cases who have recovered or died). The compartments $II(t)$ and $R(t)$ de-couple from the other compartments, and their values can be obtained from $S(t), \, E(t), \, I(t), \, C(t)$. A schematic diagram of the model is shown in figure 1. 
The system of differential equations for $S(t), \, E(t), \, I(t), \, C(t)$ is

\vspace{0.1in}

\begin{equation} \label{model}
\begin{array}{ll}
\displaystyle\dot{S}(t)= -  \underbrace{\ \ \ \  \beta S(t) \frac{I(t)}{N} \ \ \ \ }_{\substack{\text{infection rate of susceptibles}\\ \text{(homogeneous mixing)}}} \ \ \ - \ \underbrace{\ \ \ \  \epsilon S(t) \frac{C(t)}{N} \ \ \ \ }_{\substack{\text{infection rate due to improper handling} \\ \text{of deceased (homogeneous mixing)}}}\\
\\
\displaystyle\dot{E}(t)= \ \ \beta S(t)\frac{I(t)}{N} \ \ + \ \epsilon S(t) \frac{C(t)}{N} \ \  - \ \underbrace{\ \ \ \  \sigma E(t) \ \ \ \ }_{\substack{\text{rate of progression}\\ \text{to infectiousness}}} \ \ \ - \ \ \  \underbrace{\kappa (\alpha I(t)+\psi C(t)) \pi_E\omega_E \left(\frac{E(t)}{N}\right)}_{\substack{\text{removal of exposed} \\ \text{due to contact tracing}}}   \\
\\
\displaystyle\dot{I}(t)=\ \ \sigma E(t)\ \ -\ \underbrace{ \ \ \ \ \  \alpha I(t) \ \ \ \ \ }_{\substack{\text{general rate of identifying}\\ \text{and isolating infectious}}} \ \ - \ \underbrace{\gamma I(t)}_{\substack{\text{rate of}\\ \text{recovery}}} \ \ - \ \underbrace{ \ \ \  \nu I(t) \ \ \ }_{\substack{\text{rate of mortality}\\ \text{outside hospital}}} \ \ \ - \ \underbrace{\kappa (\alpha I(t)+\psi C(t)) \pi_I\omega_I \left(\frac{I(t)}{N}\right)}_{\substack{\text{rate of identifying and isolating} \\ \text{infectious due to contact tracing}}}  \\
\\
\displaystyle \dot{C}(t)=\underbrace{ \ \ \  \nu I(t) \ \ \ }_{\substack{\text{average time from symptoms}\\ \text{onset to death = 1/$\nu$}}} \ \ - \ \underbrace{ \ \ \ \psi C(t) \ \ \ }_{\substack{\text{average time until deceased}\\ \text{is properly handled}\ = \ 1 / \psi}} \\
\\
\end{array}
\end{equation}

\vspace{0.1in}

\noindent
We note that the transmission terms (involving $\beta$ and $\epsilon$) and the contact tracing terms (involving $\kappa$) are of mass-action form. We note also that the form of all the loss terms in the equations assures that the solutions remain  nonnegative for all time.

A major goal of our study is to fit the model to current reported data for Sierra Leone, Liberia, and Guinea. We note that the data available are the cumulative clinical reported cases \cite{WHO2,WHO3},  that is, the suspected cases, probable cases and confirmed cases according to  the definitions given in WHO \cite{WHO4}.  Therefore, if we denote by $CUM(t)$ the cumulative reported cases at time $ t $, then at time $ t+\Delta t$ we have 
\begin{equation}\label{cumulativecases}
CUM(t+\Delta t)= CUM(t)+ \underbrace{\ \ \ \ \   \int_t^{t+\Delta t}\alpha I(s) \,ds\ \ \ \ \ }_{\substack{\text{number of identified/isolated}\\ \text{infectious individuals}\\ \text{in time interval} \  (t,t+\Delta t)}}+\underbrace{\ \ \ \ \ \int_t^{t+\Delta t} \psi C(s) \,ds \ \ \ \ \ }_{\substack{\text{number of deceased identified}\\ \text{and properly handled in time interval} \  (t,t+\Delta t)}} 
\end{equation}
Hence, in order to fit the data we will use the above equation (\ref{cumulativecases}) with $\Delta t=1$ day and $CUM(0) =$ the initial cumulative reported cases. We note that to simplify the model, we have omitted the infection dynamics of hospital and healthcare workers. Although healthcare worker infections from patients is of great importance and requires major attention, the contribution of healthcare worker infections to new transmissions and to the cumulative reported cases  (\ref{cumulativecases}) is relatively small.

\section{The Parameters of the Model}

\begin{table}[ht]
\centering 
\begin{tabularx}{\textwidth}{>{} lX}
\toprule
Parameter/Variable & Description\\ [0.5ex]
\toprule
     $S(t)$ & The number of susceptible individuals at time $t$\\ \hline
     $E(t)$ & The number of exposed (incubating and not yet infectious) individuals at time $t$\\ \hline
		 $I(t)$ & The number of infectious individuals at time $t$\\ \hline
		 $C(t)$ & The number of deceased improperly handled at time $t$\\ \hline
		 $Q(t)$ & The number of susceptible individuals under quarantine at time $t$\\ \hline
		 $II(t)$ & The number of infectious individuals under isolation at time $t$\\ \hline
		 $R(t)$ & The number of infected recovered or properly handled deceased at time $t$\\ \hline
		 $N$ & Total national population (assumed to be constant) \\ \hline
		 $\beta$ & Transmission rate excluding improper handling of deceased  \cite{Althaus},\cite{Meltzer},\cite{Rivers} \\     \hline 
		 $\epsilon$ & Transmission rate due to improper handling of deceased  \cite{Meltzer},\cite{Rivers}\\  \hline
		 $\kappa$ & Average number of contacts traced per identified/isolated infectious individual \\   \hline
		 $1/\alpha$ & Average time from symptoms onset to identification/isolation of infectious individuals independent of contact tracing   \cite{Althaus},\cite{Legrand},\cite{Meltzer},\cite{Rivers}  \\   \hline
		 $\pi_I$  & Probability of a contact traced infectious individual is isolated without causing a new case\\   \hline
		 $\omega_I$ & Ratio of probability that contact traced individual is infectious at time of originating case identification to the probability a random individual in the population is infectious \\
\hline 
         $\pi_E$  & Probability a contact traced exposed individual is isolated without causing a new case\\   \hline
		 $\omega_E$ & Ratio of probability that contact traced individual is exposed at time of originating  case identification to the probability a random individual in the population is exposed \\ 
		 \hline
	$1/\gamma$ & Average time from symptoms onset to recovery  \cite{Althaus},\cite{Legrand},\cite{Meltzer},\cite{Rivers}  \\ 
\hline
     $1/\nu$ & Average time from symptoms onset to death \cite{Althaus},\cite{Legrand},\cite{Meltzer},\cite{Rivers}  \\ 
\hline

      $1/\sigma$  & Average incubation period  \cite{Althaus},\cite{Legrand},\cite{Meltzer},\cite{Rivers} \\       
\hline
		$1/\psi$ & Average time until improperly handled deceased is properly handled   \cite{Althaus},\cite{Meltzer},\cite{Rivers} \\ \hline			
   \bottomrule
   \end{tabularx}
    \caption{Model parameters.}\label{Table1}
   \end{table}

\vspace{0.1in}

The parameters of the model are given in Table 1.  The values of the parameters $\beta$, $\epsilon$, $\psi$ and $\alpha$ are estimated for the three countries using a least square curve fitting algorithm.  The parameters $\sigma$ and $\nu$ are taken to be values suggested by references in Table 1.  The parameter $\gamma$ is assumed to be a value such that the case mortality rate (outside hospital) is approximately $80\%$.  In fitting the model to the data, we assume that contact tracing does not occur ($\kappa=0$) during the time period of the data (up until September 23, 2014).  The reasons for this assumption are that contact tracing has been insufficient in the three countries and to reduce the number of parameters to estimate.  The main goal in incorporating contact tracing is to project forward how effective contact tracing can affect the future number of cases.

The basic reproduction number of the model (\ref{model}) is given by the following formula, computed by the next generation method \cite{van} (Appendix):
$$\mathcal R_0=\frac{\beta}{\alpha+\gamma+\nu}+\frac{\nu\epsilon}{\psi(\alpha+\gamma+\nu)}.$$
The $\mathcal R_0$ values we obtain are similar to the $\mathcal R_0$ values obtained  in
\cite{Althaus},\cite{Chowell},\cite{Fisman},\cite{Gomes},\cite{Towers},\cite{Team}. Notice that $\mathcal R_0$ does not contain the contact tracing parameters $\kappa, \, \pi_E, \, \pi_I, \, \omega_E, \, \omega_I$.  We note that  $\mathcal R_0$ is an incomplete indicator of the epidemic outcome.  Even if $\mathcal R_0 > 1$, the epidemic will ultimately be eliminated (Appendix),   and if $\mathcal R_0 < 1$, the number of cases may increase.  The value of $\mathcal R_0$ is as an approximate measure of the influence of various parameters on the number of secondary cases caused by an infectious individual, which does not account for future infections caused by these secondary cases that may be prevented by contact tracing.  Thus $\mathcal R_0$ does not provide a complete description of how the parameters affect the epidemic outcome.  

The removal (isolation) of infectious individuals due to contact tracing can be derived as follows: Let $dI(t) =$ the number of infectious individuals removed \emph{due to contact tracing} in the time interval $(t,t+dt$). Then
$$d I(t) = - \underbrace{ \ \ \ \ (\alpha I(t)+\psi C(t)) d t \ \ \ \ }_{\substack{\text{number of identified/isolated}\\ \text{infectious or deceased individuals}\\ \text{in the time interval} \  (t,t+\Delta t)}} \times \underbrace{\ \ \ \ \ \ \kappa \ \ \ \ \ \ \ }_{\substack{\text{average number of }\\ \text{contact traced individuals}\\ \text{per identified/isolated infected}}} \times \underbrace{ \ \ \ \ \pi_I \ \ \ \ }_{\substack{\text{probability of}\\ \text{compliance}}} \times \underbrace{\ \ \ \ \omega_I \frac{I(t)}{N} \ \ \ \ }_{\substack{\text{probability contact traced}\\ \text{individual is infectious}}} $$

The removal of incubating infected individuals due to contact tracing is derived similarly, and is discussed further below.
The underlying assumption is that the probability $p_E(t)$ ($p_I(t)$) a contact traced individual is incubating (infectious) at time $t$ is proportional to the probability $E(t)/N$ ($I(t)/N$) that a random individual in the population is incubating (infectious) at time $t$. The proportionality constants $\omega_E$ and $\omega_I$ are intrinsic to the infection process of family members, and others closely associated with an originating source case from which the contacts are traced. The contacts' infection likelihood is very different from the general infection likelihood of random individuals in the population. The  probabilites $p_E(t),\ p_I(t)$  can, in principle,  be ascertained from records of contact traced individuals. The probabilities $E(t)/N$ and $I(t)/N$ can be obtained from the solutions of the model. We form the ratios 
$$\omega_E = \frac{p_E(t)}{E(t)/N}, \ \ \  \omega_I = \frac{p_I(t)}{I(t)/N}$$ 
and view  $\omega_E$ and $\omega_I$ as these proportionality constants. Equivalently, we can view 
$p_E(t)$ as $\omega_E$ times as likely as $E(t)/N$ and $p_I(t)$ as $\omega_I$ times as likely as $I(t)/N$. 
The parameter $\omega_E$ may be relatively large, since a traced contact will have a much greater chance of being infected than a random individual.  For example, suppose that $p_E(t)$ is found to be $1/10$ in a given time interval and the number of incubating individuals is  $E(t)=200$.  Then, if the population is $N=4\times 10^6$, 
 
\begin{align*}
p_E(t)&= \omega_E \frac{E(t)}{N} \Rightarrow \omega_E=2000.
\end{align*}
Further, the  ratio
$$\omega_E/\omega_I \, \approx \frac{\text{average  incubation time   $= 1/\sigma$}}{\text{average time from symptoms onset to identification} =  \, 1/\alpha}$$ 
which is approximately 2 in our typical parameterizations. We note that the parameter $\psi$ could also be included in this estimation, which would decrease this ratio somewhat.

The parameters $\pi_E$ and $\pi_I$ measure the efficiency of the tracking, monitoring, and removing of incubating and infectious contacts.
In particular, $\pi_E$ measures how fast  public health workers  remove incubating individuals upon symptoms onset, and prevent secondary transmissions (so that these individuals are \emph{effectively} removed at the end of their incubation phase and transition to the $II$ compartment without causing a new case). 

We note that if $t_1$ is the time contact tracing begins, then the cumulative number of cases reported between time $0$ and time $t < t_1$ is
$$CUM(0) + \int_0^t (  \alpha I(s)  + \psi C(s) )ds,$$  and the cumulative number of cases reported at time $t > t_1$ is
$$CUM(0) +\int_0^{t_1} ( \alpha I(s) + \psi C(s) ) ds$$
$$ +  \int_{t_1}^t \Bigg(\alpha I(s) + \psi  C(s) + \kappa (\alpha I(s) + \psi  C(s)) \pi_E \omega_E \frac{E(s)}{N} + \kappa (\alpha I(s) + \psi  C(s)) \pi_I \omega_I \frac{I(s)}{N} \Bigg) ds.$$ 
Since the total cumulative number of cases, both reported and unreported, at time t is $N-S(t)$, the cumulative number of unreported cases can be determined at any time $t > 0$.

The entire contact tracing process is highly dependent on public health resources, and varies greatly in different locations and epidemic stages. For example, in S\'en\'egal the following policy has been implemented: 

\vspace{0.1in}

1) each identified  patient is questioned in order to obtain a complete list of contacts;

2) the contacts are traced;

3) each contact is asked to stay at home;

4) each day, for 21 days, a healthcare worker visits the contacts and verifies whether or not  the contacts are showing symptoms.

\vspace{0.1in}

\noindent
These protocols are rigorous and have been successful in preventing new cases in S\'en\'egal. On October 17, 2014 the World Health Organization declared the end of the outbreak of the Ebola epidemic in S\'en\'egal (after 42 days with no new cases and  with active surveillance demonstrably in place and supported by good diagnostic capacity)  \cite{WHOdeclares} . 

\section{A More General Model}

The model presented  is directly applicable to the current epidemics in Sierra Leone, Liberia, and Guinea. We discuss here a more general model that incorporates additional features of Ebola epidemics applicable to other locations. 
In order to simplify the model presented here, we have neglected several features of contact tracing in the susceptible equation, which would involve a compartment of monitored susceptibles.  There are several reasons for this simplification.  First of all, strict isolation or monitoring of contact traced individuals not showing symptoms has been problematic in West Africa during this outbreak.  Second, accounting for such isolation or monitoring would likely be more complicated than simply removing individuals from the susceptible compartment.  Indeed, contact traced susceptibles are in a higher risk group for acquiring infection, since they are likely to be in contact with exposed individuals.  For example, family members of an infected individual will continue to be in contact with each other, putting them at higher risk for a ``second-order transmission'' from another infected family member.  Such second-order effects will be considered in a more general model in future work. This more general model involves low risk and high risk susceptibles, $S_0(t)$ and $S_1(t)$, respectively, as well as an added monitored compartment $M(t)$.  Here $M(t)$ represents contact traced susceptibles sufficiently well-monitored such that they do not cause secondary infections (if infected).  Note that the compartment $M(t)$ should include health care workers who have had direct contact with infectious cases.  In addition, all contact traced susceptible individuals are assumed to be high-risk. Further, the infection rate of high-risk susceptible individuals who are not well-monitored is assumed to be a linear term in the $S_1$ equation. A schematic diagram of the more general model is shown in figure 15.  The new equations and modified transmission events are as follows:

\vspace{0.1in}

\begin{equation} \label{moregeneral}
\begin{array}{ll}
\displaystyle\dot{S}_0(t)= -  \underbrace{\beta S_0(t) \frac{I(t)}{N}}_{\substack{\text{infection rate due}\\ \text{to low-risk}\\ \text{(homogeneous mixing)}}} \ \ \ - \underbrace{\rho S_0(t) \frac{I(t)}{N}}_{\substack{\text{rate of transition}\\ \text{from low-risk}\\ \text{to high-risk}}} \ \ \ \ - \underbrace{\epsilon S_0(t) \frac{C(t)}{N}}_{\substack{\text{infection rate due to} \\ \text{ improper handling} \\ \text{ of deceased}}}  + \ \ \ \  \phi S_1(t) \ + \ \underbrace{\tau (1-\delta)M(t)}_{\substack{\text{rate of return to low-risk} \\ \text{for monitored susceptibles}}} 
\\
\\
\displaystyle\dot{S}_1(t)=\rho S_0(t) \frac{I(t)}{N} \ \ \ -  \ \ \   \underbrace{\beta S_1(t) \frac{I(t)}{N}}_{\substack{\text{infection rate due}\\ \text{to low-risk} \\ \text{(homogeneous mixing)}}} \ \ \   -  \ \ \ \underbrace{\epsilon S_1(t) \frac{C(t)}{N}}_{\substack{\text{infection rate due to} \\ \text{ improper handling} \\ \text{ of deceased}}}  \ \ -   \underbrace{\theta S_1(t)}_{\substack{\text{infection rate due to high risk} \\ \text{(not well-monitored)}}}
\\
\hspace{1.0in}
- \ \ \ \underbrace{ \ \ \  \phi S_1(t) \ \ \ }_{\substack{\text{rate of transition from} \\ \text{high-risk to low-risk}}}
- \ \ \  \underbrace{\kappa (\alpha I(t)+\psi C(t)) \pi_S\omega_S \left(\frac{S_1(t)}{N}\right)}_{\substack{\text{(effective) monitoring rate of high-risk} \\ \text{susceptibles due to contact tracing}}} \ \ \ \ + \ \underbrace{\tau \delta M(t)}_{\substack{\text{rate of return to high-risk} \\ \text{for monitored susceptibles}}} 
\\
\\
\displaystyle \dot{M}(t)= \ \ \kappa (\alpha I(t)  +  \psi C(t)) \pi_S\omega_S  \left(\frac{S_1(t)}{N}\right) \ \ \ \ \   -   \underbrace{\xi M(t)}_{\substack{\text{infection rate due to high risk} \\ \text{(well-monitored)}}}
- \ \ \ \underbrace{ \ \ \ \ \  \tau M(t) \ \ \ \  \ }_{\substack{\text{average monitoring period for}\\ \text{susceptible under surveillence}\ = \ 1 / \tau}}  \\ \\

\displaystyle\dot{E}(t)=  \beta (S_0(t)+S_1(t))\frac{I(t)}{N}  +  \epsilon (S_0(t)+S_1(t)) \frac{C(t)}{N}  + \theta S_1(t) -  \sigma E(t) -  \kappa (\alpha I(t)+\psi C(t)) \pi_E\omega_E \left(\frac{E(t)}{N}\right)    \\
\\
\end{array}
\end{equation}

\vspace{0.1in}

\begin{table}[h]
\centering 
\begin{tabularx}{\textwidth}{>{} lX}
\toprule
Parameter/Variable & Description\\ [0.5ex]
\toprule
                    $S_0(t)$ & The number of low risk susceptible individuals at time $t$\\ \hline
                    $S_1(t)$ & The number of high risk susceptible individuals at time $t$\\ \hline
                    $M(t)$ & The number of susceptible individuals under effective monitoring at time $t$\\ \hline
		 $\rho$ & Transition rate from low risk to high risk  \\     \hline 
		 $\phi$ & Transition rate from high risk to low risk \\    \hline
		 $1/\tau$ & Average monitoring period for contact traced susceptible individuals \\   \hline		
                    $\delta$ & Fraction of monitored individuals returned to high risk susceptible\\   \hline
		 $\theta$ & Infection rate due to high risk (of individuals not well-monitored)\\		 \hline
		 $\pi_S$  & Probability a contact traced susceptible individual does not cause secondary cases (if infected while monitored)  \\   \hline	
		 $\omega_S$ & Ratio of the probability a contact traced individual is a high-risk susceptible to the probability a random individual in the population
		       is susceptible \\ \hline
		       $\xi$ & Infection rate due to high risk of  well-monitored individuals (second order infections)\\
   \bottomrule
   \end{tabularx}
    \caption{More general model parameters.}\label{Table2}
   \end{table}
\noindent 
The $I(t)$ and $C(t)$ equations are the same as in (\ref{model}).  Also, the loss term $\xi M(t)$ transitions the infected (monitored) individuals into the $II(t)$ compartment, which is decoupled from the equations in (\ref{moregeneral}).   Note that these individuals would be removed upon symptoms onset, and because they are well-monitored, they can be considered \emph{effectively} removed in this transition term. 

It is possible to generalize the model further  by tracking the disease stage of all infected individuals, particularly contact traced individuals.  Such a model is best treated with an age of infection variable, which allows tracking of the incubation and infectious disease stages \cite{Huo}. In future work we will develop these models, with age of infection as a continuum independent variable. 

\section{Simulations of the Ebola Epidemic in Sierra Leone} 

In figure 2 we fit the model without contact tracing to the cumulative reported case data for Sierra Leone from May 27, 2014 to September 23, 2014 (WHO \cite{WHO2,WHO3}).  The fit to data can be accomplished with varying combinations of parameters. Here we have used a least-squares algorithm to obtain a  choice of parameters with relatively accurate fit. The parameters obtained in the fit yield a basic reproduction number of 
$\mathcal R_0=1.26$. The simulation yields the following information about the epidemic  on September 23, 2014 (day 119): the ratio of exposed cases to infectious cases is $E(119) / I(119) \approx 2.49$; the ratio of improperly handled deceased cases to infectious cases is $C(119) / I(119) \approx 0.57$; and the ratio of cumulative  reported cases to cumulative unreported cases is  $\approx 1.78$. These ratios, which are dependent on parameters, are relatively stable at the data end-stage. In figure 3 we add contact tracing to the model and predict the further evolution of the epidemic in Sierra Leone forward from September 23, 2014. The contact tracing parameters $\alpha$ and $\kappa$ are varied in a sensitivity analysis, while the other contact tracing parameters are held constant. The graphics reveal that a general identification/isolation rate $\alpha > 0.3$ is required for containing the epidemic. The number $\kappa$ of contacts traced per identified case is also important if $\alpha$ is smaller. After contact tracing begins and for a short time, the reported cases increase as $\kappa$ increases, but then the epidemic subsides as contact tracing takes effect, and the reported cases decrease as $\kappa$ increases.

\section{Simulations of the Ebola Epidemic in Liberia} 

In figure 4 we fit the model without contact tracing to the cumulative reported case data for Liberia from June 17, 2014 to September 23, 2014 (WHO  \cite{WHO2,WHO3}).  We have used a least-squares algorithm to obtain a  choice of parameters with relatively good fit (other parameter choices will give similar fits). The parameters obtained in the fit yield a basic reproduction number of $\mathcal R_0=1.54$. The simulation yields the following information about the epidemic  on September 23, 2014 (day 98): the ratio of exposed cases to infectious cases is $E(98) / I(98) \approx 3.35$; the ratio of improperly handled deceased cases to infectious cases is $C(98) / I(98) \approx 0.58$; and the ratio of cumulative reported cases to cumulative unreported cases is  $\approx 1.37$. These ratios, again depend on parameters, are very stable throughout most of the period of simulation. In figure 5 we add contact tracing to the model and predict the further evolution of the epidemic in Liberia forward from September 23, 2014. The contact tracing parameters $\alpha$ and $\pi_E$ are varied in a sensitivity analysis, while all  the other contact tracing parameters are held constant. The graphics again reveal that an identification/isolation rate $\alpha > 0.3$ is required for containing the epidemic. The role of $\pi_E$, as the probability of efficiently tracing and monitoring an incubating contacted individual, who is infected, but not yet infectious, is also important for the containment of the epidemic. As in figure 3,  the reported cases increase for a short time as $\pi_E$ increases, but then the  decrease as $\pi_E$ increases, as contact tracing takes effect. In figure 8 we illustrate the nonlinear effect of varying $\pi_E$ against the cumulative  projected cases  forward 100 days.

\section{Simulations of the Ebola Epidemic in Guinea} 

In figure 6 we fit the model without contact tracing to the cumulative reported case data for Guinea from March 25, 2014 to September 23, 2014 (WHO  \cite{WHO2,WHO3}).  We have used a least-squares algorithm to optimize a  choice of parameters for the fitting (other choices will yield a similar fit). The fit is problematic in the middle range of the time period, but the data is very erratic, and the cumulative count even decreases some days. The parameters obtained in the fit yield a basic reproduction number of $\mathcal R_0=1.11$. The simulation yields the following information about the epidemic  on September 23, 2014 (day 182): the ratio of exposed cases to infectious cases is $E(182) / I(182) \approx 2.56$; the ratio of improperly handled deceased cases to infectious cases is $C(182) / I(182) \approx 0.66$; and the ratio of cumulative reported cases to cumulative unreported cases is  $\approx 4.95$. These ratios, again depend on parameters, and again are very stable throughout most of the period of simulation. In figure 7 we add contact tracing to the model and predict the further evolution of the epidemic in Guinea forward from September 23, 2014. The contact tracing parameters $\alpha$ and $\pi_I$ are varied in a sensitivity analysis, while all the other contact tracing parameters are held constant. The graphics again reveal that the general identification/isolation rate $\alpha > 0.3$ is of great importance in containing the epidemic. The efficiency of the contact tracing parameter $\pi_I$ is of lesser importance in these simulations than the parameter $\pi_E$, especially when $\alpha > 0.3$ (compare to figure 5). The probability $\pi_I$ of efficiently tracing and isolating an infectious contacted individual is compensated by the general efficiency of isolating and removing an infectious individual (independent of contact tracing). The probability $\pi_E$ of identifying and isolating an incubating contacted individual so that they cause no new infections is dependent  solely on the contact tracing process and its efficiency in monitoring and isolation of these incubating individuals upon symptoms onset. After contact tracing begins, the cumulative reported  cases and cumulative total  cases decrease as 
$\pi_I$ increases without the switch-over observed in figures 3 and 5.
  
\section{A Stochastic Version of the Model}  
  
Figures 9,10,11,12,13, and 14 show 100 stochastic simulations compared with the ODE solution for contact tracing in Sierra Leone with different rates of case identification $\alpha$.  The stochastic simulations are generated by simulating a Continuous Time Markov Chain as a continuation of the ODE solution beginning at the last data time point.  The contact tracing parameters are assumed to be $\omega_E=2000, \, \omega_I=1000$, $\pi_E=0.5, \, \pi_I=0.8$ and $\kappa=20$ in Figures 9, 10, and 11.  In Figures 12 and 13, we assume $\kappa=0$ (no contact tracing), while $\kappa=10$, $\pi_E=0.1, \, \pi_I=0.5$ (less effective contact tracing) in Figure 14. All other parameters are taken to be the same as in the previous fit to the Sierra Leone epidemic.
The averages of the stochastic model solutions agree with the ODE solutions  of (\ref{model}).

\section{Summary and Conclusions} 

We have developed a model of Ebola epidemics in West Africa that focuses attention on the elements of public health policies for containment of these epidemics. Our simulations for Sierra Leone, Liberia, and Guinea fit the cumulative reported cases for these countries up to September 23, 2014, and project future epidemic levels forward from September 23, 2014 (based on various parameterizations corresponding to these elements). Our projections indicate that the most important elements for containment of the epidemics  within a relatively short time span are that

\vspace{.1in}

(1) infectious cases (independent of contact tracing) are efficiently reported and isolated, with the average time between the appearance of symptoms and isolation  less than 3 days ($\alpha > 0.3$);

\vspace{.1in}

(2) contact traced incubating infected cases are efficiently monitored, with average probability of compliance, with isolation upon appearance of symptoms (such that no new cases are caused by individual), greater than 0.5 ($\pi_E > 0.5$).

\vspace{.1in}

\noindent
Also of importance in mitigation of the epidemics is a reduced rate at which infected deceased are improperly handled ($\psi, \, \nu$), a sufficient number of contacts traced per identified infectious individual ($\kappa$), and an efficient identification and isolation of contact traced infectious individuals ($\pi_I$). The model allows quantification of  the parameters corresponding to public health controls ($\alpha, \, \psi,  \, \kappa, \, \pi_E, \, \pi_I, \, \omega_E, \, \omega_I$) for evaluating the impact of public health policies for the evolution of these epidemics.

\newpage
\normalsize

\appendix
\section*{Appendix}

\textbf{Theorem.}
Let $\beta, \sigma, \alpha, \gamma, \nu, \psi > 0$,  and $\epsilon, \kappa, \pi_E, \pi_I, \omega_E, \omega_I \geq 0$. There exists a unique solution $S(t)$, $E(t)$, $I(t)$, $C(t)$ to \eqref{model} for $t \geq 0$ and initial values $S(0) = S_0 > 0$, $E(0) = E_0 \geq 0$, $I(0) = I_0 \geq 0$, $C(0) = C_0 \geq 0$, and the solution satisfies $S(t) \geq 0, E(t) \geq 0, I(t) \geq 0, C(t) \geq 0$ for all $t \geq 0$. All steady states of  \eqref{model} in the positive cone of $\R^4$ have the form $(\bar S,0,0,0)$, where  $\bar S \geq 0$. Further,
$$\int_0^{\infty} E(t) dt \leq \frac{S(0) + E(0)}{\sigma},\, \int_0^{\infty} I(t) dt \leq \frac{S(0) + E(0) + I(0)}{\alpha + \gamma + \nu},$$
$$\int_0^{\infty} C(t) dt \leq \frac{C(0) (\alpha + \gamma + \nu) + \nu (S(0) + E(0) + I(0))}{\psi (\alpha + \gamma + \nu)}.$$ The solutions have the following asymptotic behavior: $S(t)$ is nonincreasing,
$$ \lim_{t \rightarrow \infty} S(t) = S(0) \exp\Bigg[ - \frac{\beta}{N} \int_0^{\infty} I(t) dt - \frac{\epsilon}{N}  \int_0^{\infty} C(t) dt\Bigg] \, \,
= \, \, S(0)  \exp\Bigg[ - \frac{\epsilon \, C(0)}{N \psi} - \frac{\beta \psi + \epsilon \nu}{N \psi} \int_0^{\infty} I(t) dt \Bigg] >  0,$$ 
$$\text{ and } \lim_{t \rightarrow \infty} E(t) = 0, \, \lim_{t \rightarrow \infty} I(t) = 0, \, \lim_{t \rightarrow \infty} C(t) = 0.$$

\vspace{0.1in}

\noindent
\textbf{Proof.} 
The system (\ref{model}) has locally Lipschitz continuous nonlinear terms and is quasi-positive. Therefore, a unique solution $S(t), E(t), I(t), C(t)$ is defined and nonnegative on a maximal interval of existence for nonnegative initial values. For $t$ in the maximal interval of existence, 
$$S^{\prime}(t) + E^{\prime}(t) + I^{\prime}(t) \leq - (\alpha + \gamma + \nu) I(t) \, \, \Rightarrow$$
$$S(t) + E(t) + I(t) + (\alpha + \gamma + \nu) \int_0^t I(s) ds  \,  \leq \, S(0) + E(0) + I(0) \, \, \Rightarrow \, \, 
\int_0^t I(s) ds  \,  \leq \, \frac{S(0) + E(0) + I(0}{\alpha + \gamma + \nu}.$$
Further, 
$$S^{\prime}(t) + E^{\prime}(t) \leq - \sigma E(t) \, \, \Rightarrow \, \, S(t) + E(t) + \sigma \int_0^t E(s) ds \leq S(0) + E(0) \Rightarrow 
\int_0^t E(s) ds \leq \frac{S(0) + E(0)}{\sigma}.$$
Also, 
$$C^{\prime}(t) + \psi C(t) = - \nu I(t) \, \, \Rightarrow \, \, C(t) + \psi \int_0^t C(s) ds = C(0) +  \nu \int_0^t I(s) ds   \Rightarrow$$  
$$C(t) \leq C(0) + \nu \Bigg( \frac{S(0)+E(0)+I(0)}{\alpha + \gamma + \nu}\Bigg) \, \text{and} \,  \int_0^t C(s) ds  \leq \frac{C(0)( \alpha + \gamma + \nu)+\nu(S(0) + E(0) + I(0))}{\psi (\alpha + \gamma + \nu)}.$$
Thus, $S(t)$, $E(t)$, $I(t)$, $C(t)$ stay bounded in the positive cone of $\R^4$ on every bounded interval in $[0,\infty)$, and the maximal interval of existence must be $[0,\infty)$. 

\vspace{0.1in}

Let $(\bar{S},\bar{E},\bar{I},\bar{C})$ be a steady state in the positive cone of $\R^4$. From the  first equation in (\ref{model}) 
$$\beta \bar{S} \bar{I} \, + \, \epsilon \bar{S} \bar{C} = 0 \Rightarrow \bar{S} = 0  \text{ or } \bar{S} > 0 \text{ and } \bar{I} = \bar{C} =0.$$
From the second equation in (\ref{model})
$$\beta \bar{S} \bar{I} \, + \, \epsilon \bar{S} \bar{C}\, = \, \sigma \bar{E}  \, + \, \kappa (\alpha \bar{I} + \psi \bar{C}) \pi_E \omega_E \frac{\bar{E}}{N} \, \Rightarrow \, \bar{E} = 0.$$
From the third equation in (\ref{model}) $\bar{I} = 0$ and from the fourth equation in (\ref{model}) $\bar{C} = 0$.
Since $E(t), I(t), C(t)$ are nonnegative for $t \geq 0$, $E^{\prime}(t)$, $I^{\prime}(t)$, $C^{\prime}(t)$ are bounded for $t \geq 0$,
and $\int_0^{\infty}E(t) dt < \infty$, $\int_0^{\infty} I(t) dt  < \infty$, $\int_0^{\infty} C(t) dt < \infty$, it follows that $\lim_{t \rightarrow \infty} E(t) = 0$, $\lim_{t \rightarrow \infty} I(t) =  0$, $\lim_{t \rightarrow \infty} C(t) = 0$. Then, from the formula for $C(t)$ above
$$\int_0^{\infty} C(t) dt = \frac{C(0) + \nu \int_0^{\infty} I(t) dt}{\psi}.$$
From the fourth equation in (\ref{model}) 
$$C(t) =e^{-\psi t} C(0) \ + \, \nu \int_0^t e^{-\psi(t-s)} I(s) ds.$$
From the first equation in (\ref{model}) 
\begin{equation} \label{appendix4}
\begin{array}{ll}
S(t) = S(0) \exp \Bigg[ - {\displaystyle \int_0^t} \Bigg( \frac{\beta}{N} I(s) + \frac{\epsilon}{N} C(s) \Bigg)ds \Bigg]\\
\\
= S(0) \exp \Bigg[ - {\displaystyle \int_0^t}
\Bigg( \frac{\beta}{N} I(s) + \frac{\epsilon}{N} 
(e^{-\psi s} C(0) \ + \, \nu {\displaystyle \int_0^s e^{-\psi(s-r)}} I(r) dr ) \Bigg) ds \Bigg]\\
\\
= S(0) \exp \Bigg[ - \frac{\beta}{N} {\displaystyle \int_0^t} I(s) ds \Bigg] \exp \Bigg[ \frac{- \epsilon \,  C(0)}{N \psi} (1-e^{-\psi t}) \Bigg] 
\exp \Bigg[ -\frac{\epsilon \, \nu}{N} {\displaystyle \int_0^t e^{\psi r}} I(r) \Bigg( {\displaystyle \int_r^t e^{-\psi s}} ds \Bigg) dr \Bigg]\\
\\
= S(0) \exp \Bigg[ - \frac{\beta}{N} {\displaystyle \int_0^t}I(s) ds \Bigg] \exp \Bigg[ \frac{- \epsilon \,  C(0)}{N \psi} (1-e^{-\psi t}) \Bigg] 
\exp \Bigg[ -\frac{\epsilon \, \nu}{N \psi} {\displaystyle \int_0^t} I(r) dr \Bigg]\ 
\exp \Bigg[ -\frac{\epsilon \, \nu}{N \psi \, e^{\psi t}} {\displaystyle \int_0^t} e^{\psi r} I(r) dr  \Bigg]\
\end{array}
\end{equation}
Observe that
$$\lim_{t \rightarrow \infty} \frac{{\displaystyle \int_0^t} e^{\psi r} I(r) dr}{ e^{\psi t} }
= 0 \text{ if } {\displaystyle \int_0^{\infty}} e^{\psi r} I(r) dr < \infty$$ 
and, since $\lim_{t \rightarrow \infty} I(t) = 0$,
$$\lim_{t \rightarrow \infty} \frac{{\displaystyle \int_0^t} e^{\psi r} I(r) dr}{ e^{\psi t} }
= \lim_{t \rightarrow \infty} \frac{e^{\psi t}I(t)}{\psi e^{\psi t}}
= 0 \text{ if } {\displaystyle \int_0^{\infty}} e^{\psi r} I(r) dr =\infty.$$
Thus, (\ref{appendix4}) implies 
$$\lim_{t \rightarrow \infty} S(t) = S(0) \exp\Bigg[ - \frac{\beta}{N} \int_0^{\infty} I(t) dt - \frac{\epsilon}{N}  \int_0^{\infty} C(t) dt\Bigg] \, \,
= \, \, S(0)  \exp\Bigg[ - \frac{\epsilon \, C(0)}{N \psi} - \frac{\beta \psi + \epsilon \nu}{N \psi} \int_0^{\infty} I(t) dt\Bigg] >  0.$$
\hfill $\square$

\vspace{0.2in}

\noindent
\textbf{Remark.}
We provide here an explanation of the computation of $\mathcal R_0$. Let $S = N$ and let 
$$F_1(E,I,C) =  \beta I + \epsilon C-  \sigma E-\kappa (\alpha I+\psi C) \pi_E\omega_E \left(\frac{E}{N}\right),$$ 
$$F_2(E,I,C) = \sigma E - (\alpha + \gamma + \nu) I-\kappa (\alpha I+\psi C) \pi_I\omega_I \left(\frac{I}{N}\right),$$
$$F_3(E,I,C) = \nu I - \psi C.$$
The Jacobian of $(F_1,F_2,F_3)$ at $(0,0,0)$ is 
$$J(0,0,0) = \begin{bmatrix} -\sigma & \beta & \epsilon \\ \sigma & -(\alpha + \gamma + \nu)  & 0 \\ 0 &  \nu & - \psi \end{bmatrix} = 
\begin{bmatrix} 0 & \beta & \epsilon \\ 0 & 0  & 0 \\ 0 &  0 &  0\end{bmatrix}
-  \begin{bmatrix} \sigma & 0 & 0 \\  - \sigma & \alpha + \gamma + \nu  & 0 \\ 0 & -  \nu &  \psi \end{bmatrix} = F - V.$$
The formulas for $V ^{-1}$ and  $F \, V^{-1}$ are 
$$V^{-1} =  \begin{bmatrix} 1/\sigma & 0 & 0 \\  1 / (\alpha + \gamma + \nu) & 1 / (\alpha + \gamma + \nu)& 0 \\ 
\nu /  \psi (\alpha + \gamma + \nu) & \nu /  \psi (\alpha + \gamma + \nu) &1 / \psi \end{bmatrix}, \ \ \ \ 
F \, V^{-1} = \begin{bmatrix} \frac{\beta \psi + \nu \epsilon}{\psi(\alpha+\gamma+\nu)} & \frac{\beta \psi + \nu \epsilon}{\psi(\alpha+\gamma+\nu)} & \frac{\epsilon}{\psi} \\  0 & 0  & 0 \\ 0 & 0 & 0\end{bmatrix}.$$ 
The eigenvalues of  $F \, V^{-1}$ are $0$, $0$, and $(\beta \psi + \nu \epsilon) / (\psi (\alpha+\gamma+\nu)),$
and the spectral radius of $F \, V^{-1}$ is
$\mathcal R_0 = (\beta \psi + \nu \epsilon) / (\psi (\alpha+\gamma+\nu)).$

\hfill $\square$

\newpage
\begin{figure}[ht]
\begin{center}
 \includegraphics[width=6.5in,height=5.0in]{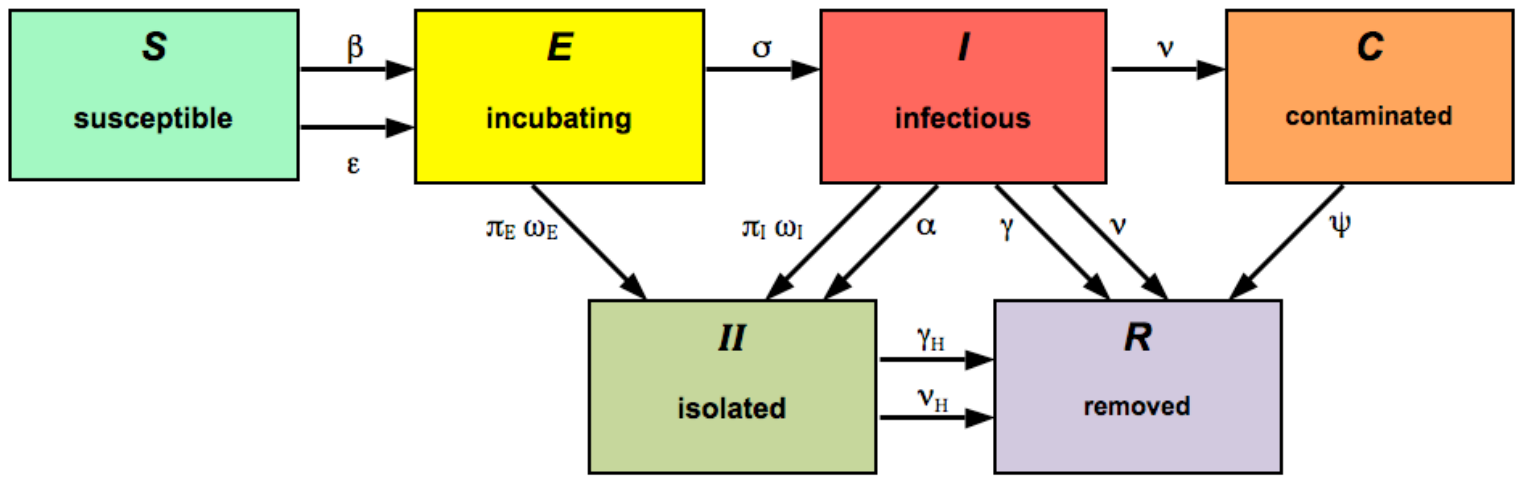}\\
\caption{Schematic diagram of the model compartments and parameters.}\label{Fig1}
\end{center}
\end{figure}

\begin{figure}[htp]
\begin{center}
\includegraphics[width=4.5in,height=2.8in]{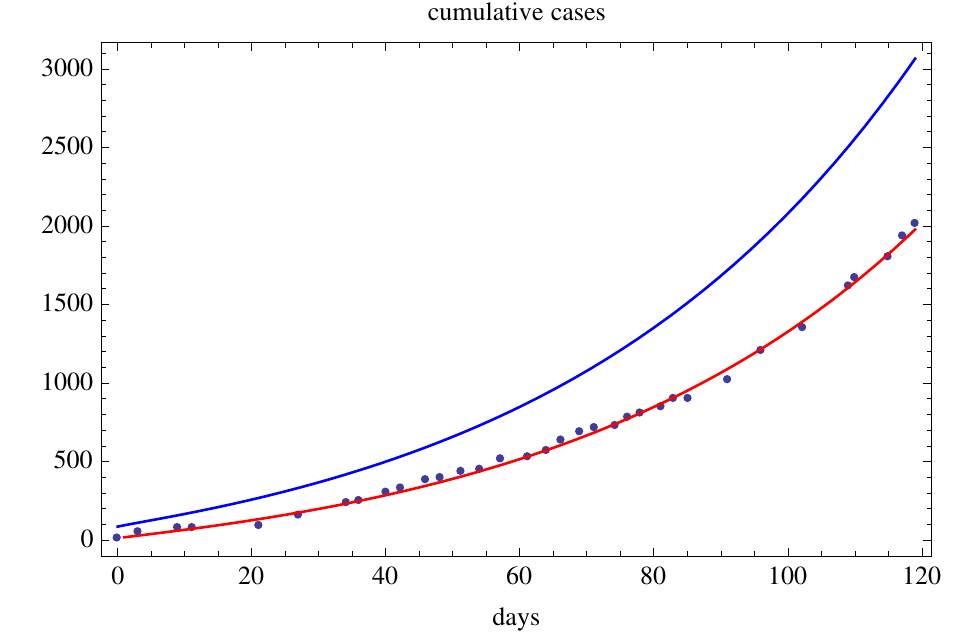}\\
\caption{Simulation of cumulative cases in Sierra Leone from May 27, 2014 to September 23, 2014. The black dots are data of cumulative reported cases (WHO), the red graph is the model simulation of cumulative reported cases, and the blue graph is the model simulation of cumulative total cases.
The parameter values are $N=6,000,000$, $\beta = 0.32$, $\epsilon = 0.0078$, $\psi=0.2$, $\alpha=0.1$,$\sigma = 1/9$, $\gamma = 1/30$, $\nu = 1/8$, the initial conditions are $S(0)=N$, $E(0)=47$, $I(0)=26$, $C(0)=12$. $\mathcal R_0=1.26.$}\label{Fig2}
\end{center}
\end{figure}

\begin{figure}[htp]
\begin{center}
\includegraphics[width=6.5in,height=4.0in]{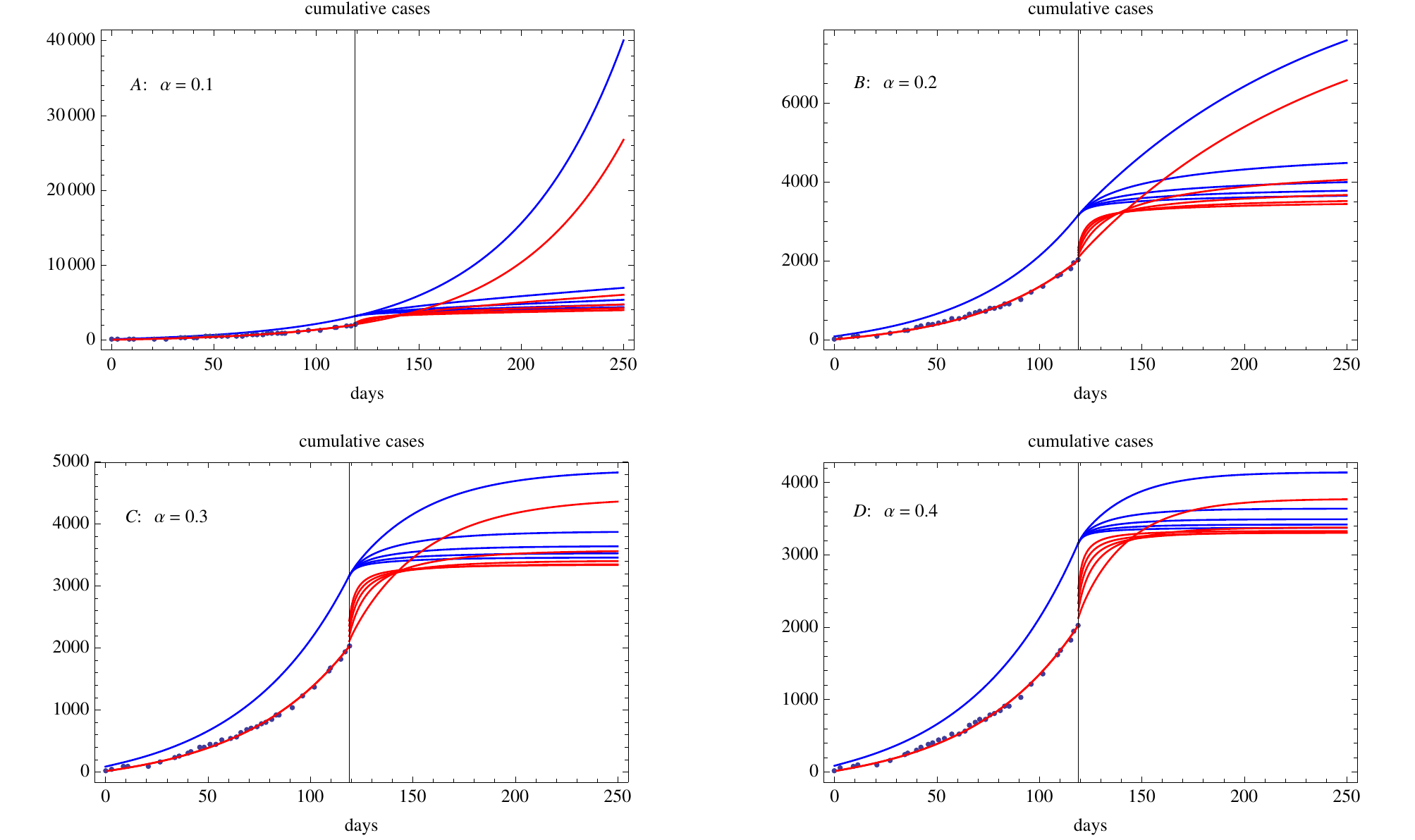}\\
\caption{Simulation of predicted cumulative cases in Sierra Leone forward from September 23, 2014 (vertical line). The red graphs are model simulations of reported cases with $\kappa= 0, \, 10, \,  20, \, 30, \, 40$.  The blue graphs are model simulations of total cases with $\kappa = 0, \, 10, \, 20, \, 30, \, 40$ (in descending order). A: $\alpha = 0.1$. B: $\alpha = 0.2$. C: $\alpha = 0.3$. D: $\alpha = 0.4$. The parameter values are is in Figure 2 and $\pi_E = 0.5$,  $\pi_I = 0.8$, $\omega_E = 2000$,  $\omega_I = 1000$. The initial conditions at September 23, 2014 are $S(119) = 5,996,940$, $E(119) = 475$, $I(119) = 191$, $C(119) = 109.$}\label{Fig3}
\end{center}
\end{figure}

\begin{figure}[htp]
\begin{center}
\includegraphics[width=4.5in,height=2.8in]{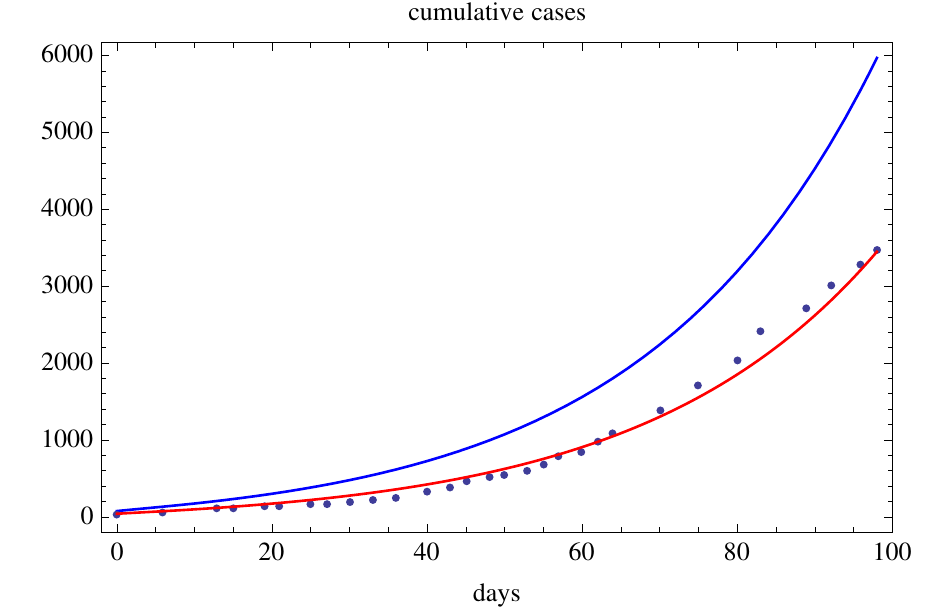}\\
\caption{Simulation of cumulative cases in Liberia from June 17, 2014 to September 23, 2014. The black dots are data of cumulative reported cases (WHO), the red graph is the model simulation of cumulative reported cases, and the blue graph is the model simulation of cumulative total cases. The parameter values are $N=4,000,000$, $\beta = 0.3$, $\epsilon = 0.316$, $\psi=0.18$, $\alpha=0.18$, $\sigma = 1/9$, $\gamma = 1/30$, $\nu = 1/8$, the initial conditions are $S(0)=N$, $E(0)=40$, $I(0)=22$, $C(0)=12$.  $\mathcal R_0=1.54$.}\label{Fig4}
\end{center}
\end{figure}

\begin{figure}[htp]
\begin{center}
\includegraphics[width=6.5in,height=4.0in]{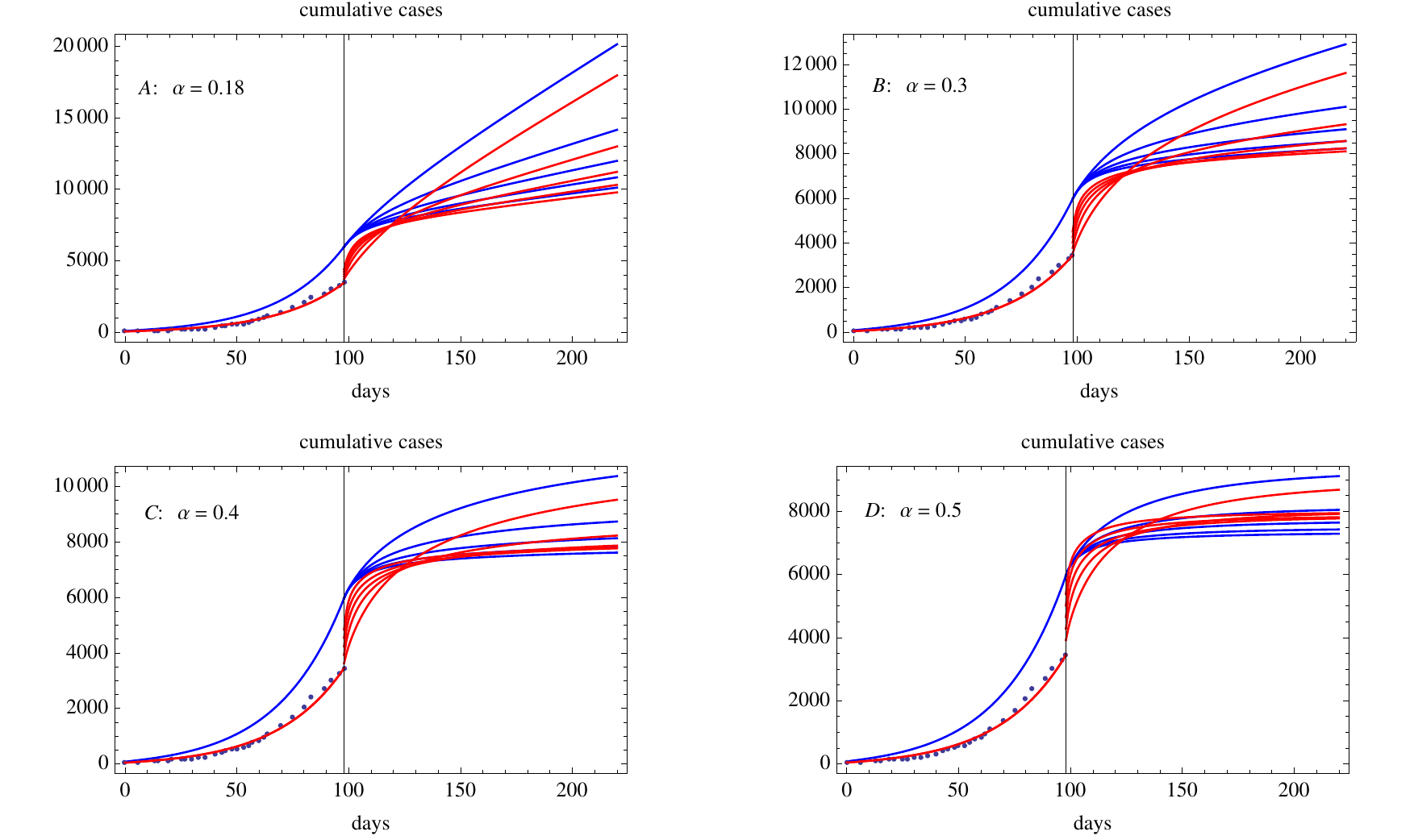}\\
\caption{Simulation of predicted cumulative cases in Liberia forward from September 23, 2014 (vertical line). The red graphs are model simulations of reported cases with $\pi_E= 0.1, \,0.3, \,  0.5, \, 0.7, \, 0.9$  The blue graphs are model simulations of total cases with $\pi_E= 0.1, \,0.3, \, 0.5, \, 0.7, \, 0.9$ (in descending order). A: $\alpha = 0.2$. B: $\alpha = 0.3$. C: $\alpha = 0.4$. D: $\alpha = 0.5$. The parameter values are is in Figure 4 and $\kappa = 20$,  $\pi_I = 0.5$, $\omega_E = 1000$,  $\omega_I = 500$. The initial conditions at September 23, 2014 are $S(92) = 3,994,061$, $E(92) = 1544$, $I(92) = 433$, $C(92) = 101.$}\label{Fig5}
\end{center}
\end{figure}

 \begin{figure}[htp]
\begin{center}
   \includegraphics[width=4.5in,height=2.8in]{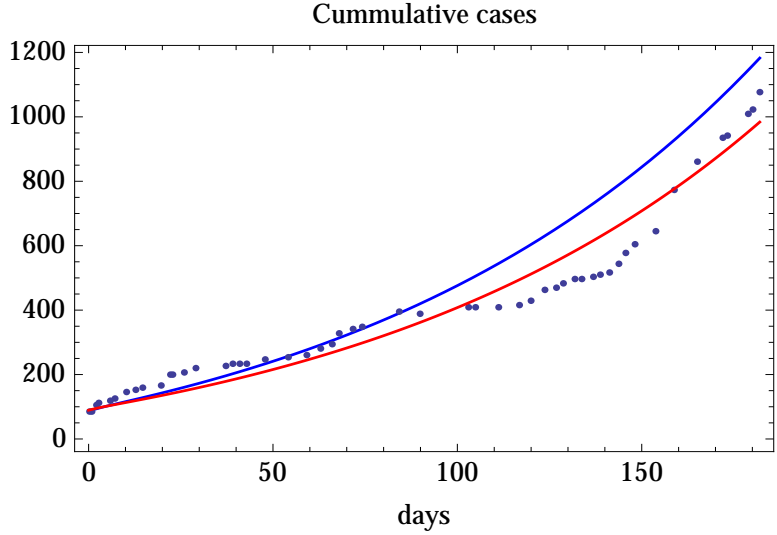}\\
  \caption{Simulation of cumulative cases in Guinea from March 25 to September 23, 2014. The black dots are data of cumulative reported cases (WHO), the red graph is the model simulation of cumulative reported cases, and the blue graph is the model simulation of cumulative total cases.
 The parameter values are $N=12,000,000$, $\beta = 0.24$, $\epsilon = 0.224$, $\psi=0.18$, $\alpha=0.2$, $\sigma = 1/7$, $\gamma = 1/32$, $\nu = 1/8$, the initial conditions are $S(0)=N$, $E(0)=6$, $I(0)=3$, $C(0)=15$. $\mathcal R_0=1.12$.}\label{Fig6}
  \end{center}
  \end{figure}

\begin{figure}[htp]
\begin{center}
   \includegraphics[width=6.5in,height=4.0in]{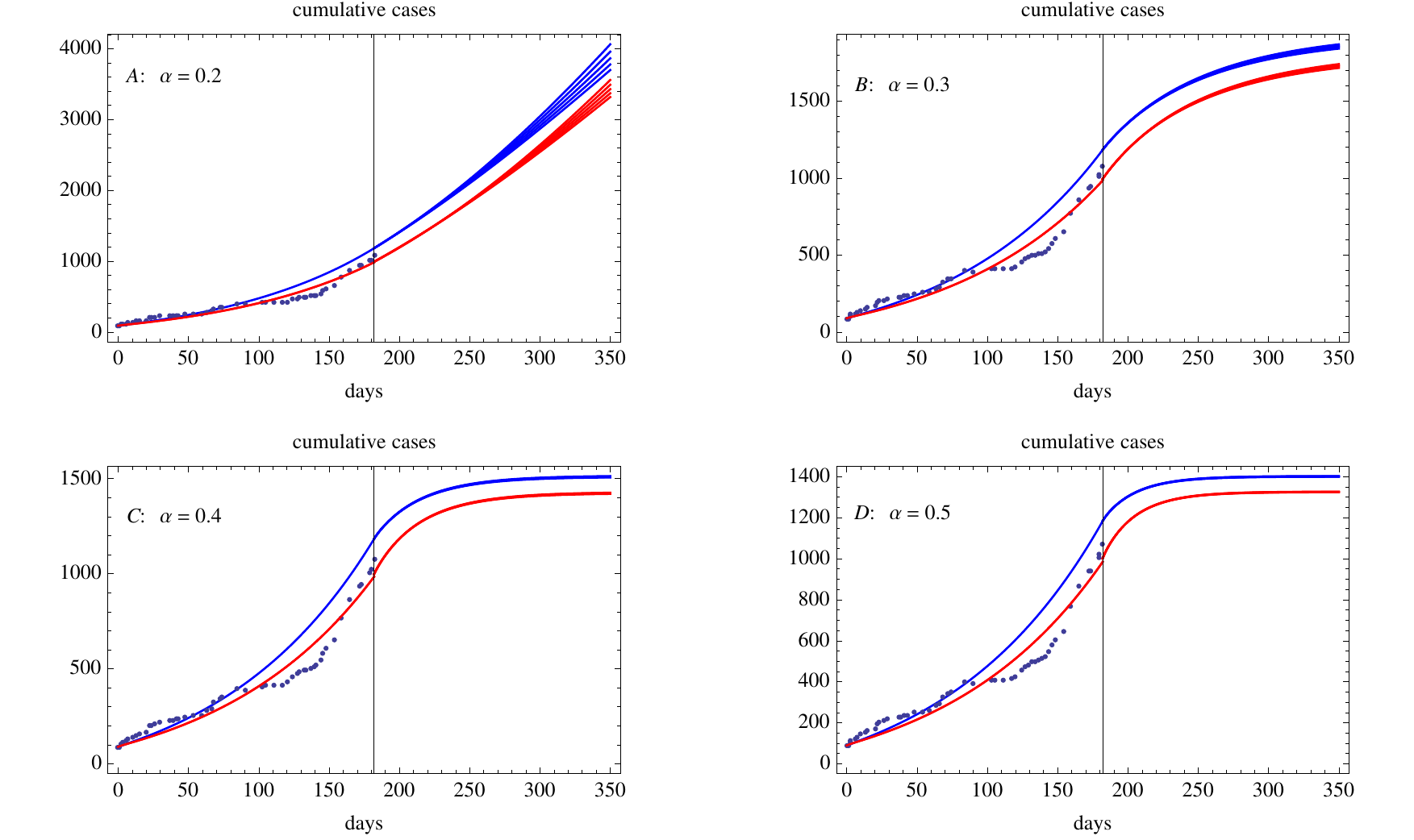}\\
  \caption{Simulation of predicted cumulative cases in Guinea forward from September 23, 2014 (vertical line). The red graphs are model simulations of reported cases with $\pi_I= 0.1, \,0.3, \,  0.5, \, 0.7, \, 0.9$ (in descending order). The blue graphs are model simulations of total cases with $\pi_I = 0.1, \,0.3, \, 0.5, \, 0.7, \, 0.9$ (in descending order). A: $\alpha = 0.2$. B: $\alpha = 0.3$. C: $\alpha = 0.4$. D: $\alpha = 0.5$. The parameter values are is in Figure 6 and $\kappa = 10$,  $\pi_E = 0.4$, $\omega_E = 2000$,  $\omega_I = 1000$. The initial conditions at September 23, 2014 are $S(182) = 11,998,903$, $E(182) = 80$, $I(182) = 31$, $C(182) = 21.$}\label{Fig7}
  \end{center}
  \end{figure}

 \begin{figure}[!ht]
 \centering
  \subfigure[][]{\label{Fig8a}\includegraphics[width=6.05in,height=2.0in]{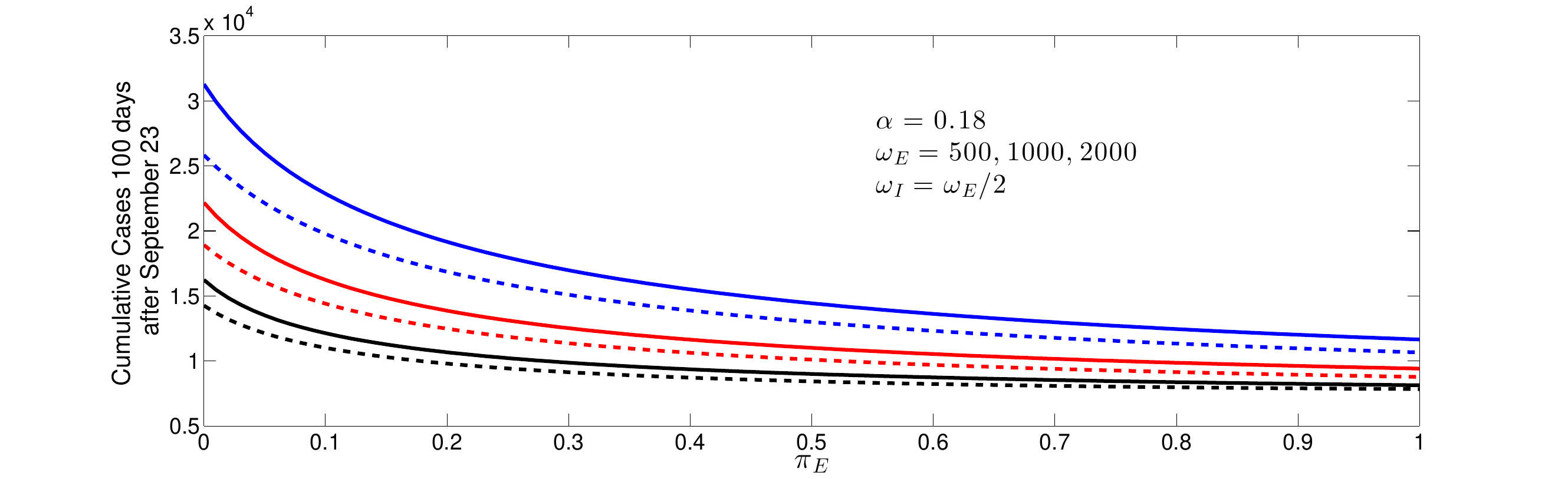}}\\
  \subfigure[][]{\label{Fig8b}\includegraphics[width=6.0in,height=2.0in]{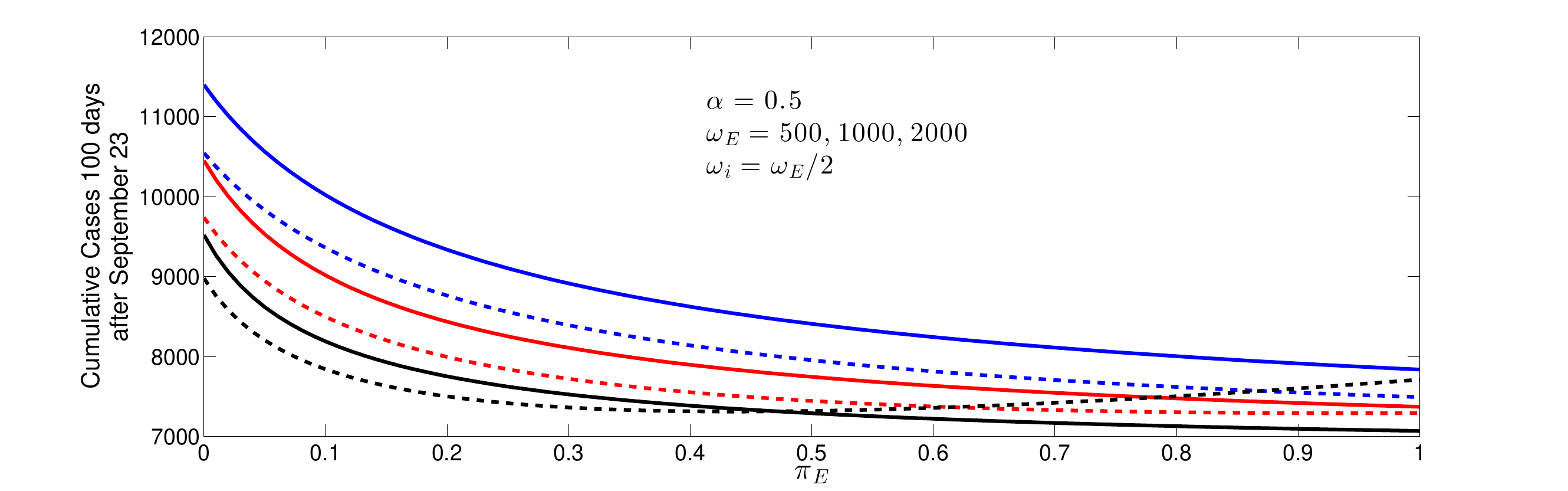}} 
    \caption{Predicted cumulative cases in Liberia 100 days forward from September 23, 2014 versus $\pi_E$ with $\omega_E=500$ (blue), $\omega_E=1000$ (red), $\omega_E=2000$ (black) and $\omega_I=\omega_E$.  The other contact tracing parameters are $\kappa=20,\pi_I=0.5$.  The case identification rate $\alpha$ is varied: (a) $\alpha=0.18$, (b) $\alpha=0.5$. The remaining parameters are as in the fit for Liberia.  The solid lines represent total cases, while the dotted lines represent reported cases.  }  
  \end{figure}

\begin{figure}[!ht]
\subfigure[][]{\label{RepAlp10}\includegraphics[width=3.25in,height=2.0in]{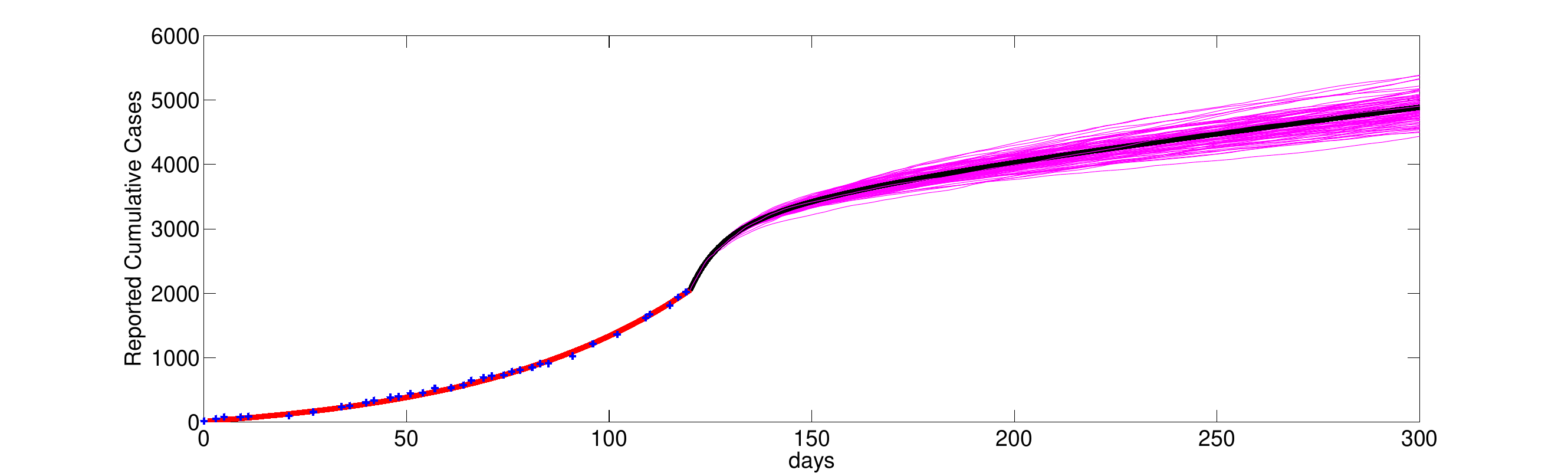}}
\subfigure[][]{\label{TotAlp10}\includegraphics[width=3.25in,height=2.0in]{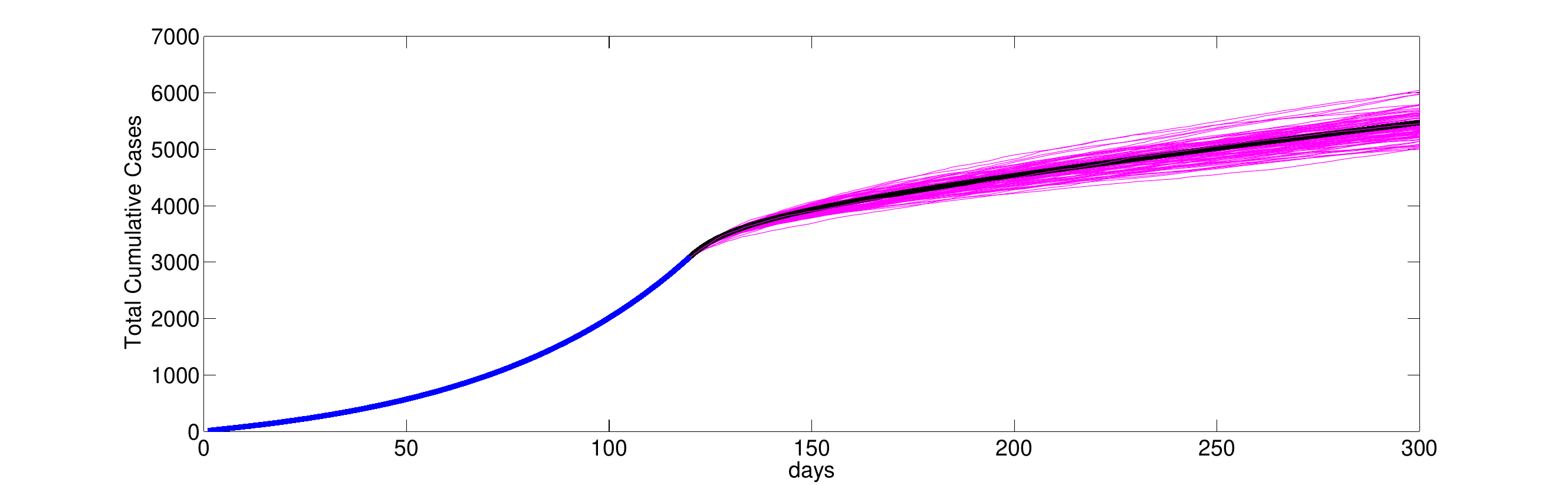}} 
  \caption{(a)  100 stochastic simulations of predicted cumulative reported cases in Sierra Leone forward from September 23, 2014, with $\alpha=0.1$, $\kappa=20$, $\pi_E=0.5$, $\pi_I=0.8$, $\omega_E=2000$, $\omega_I=1000$. The stochastic simulations are in magenta and the ODE solution is in black.  (b) Simulations of predicted total cases corresponding to (a). }
    \label{fig:mig}
\end{figure}

\begin{figure}[!ht]
\subfigure[][]{\label{RepAlp11}\includegraphics[width=3.25in,height=2.0in]{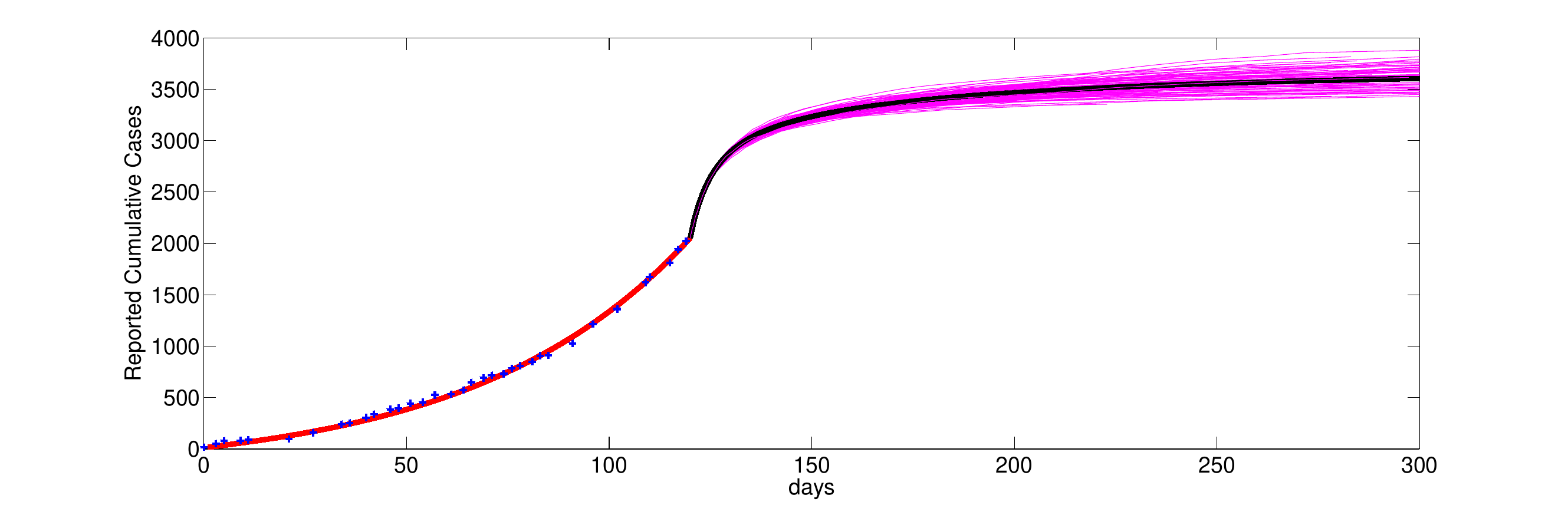}}
\subfigure[][]{\label{TotAlp11}\includegraphics[width=3.25in,height=2.0in]{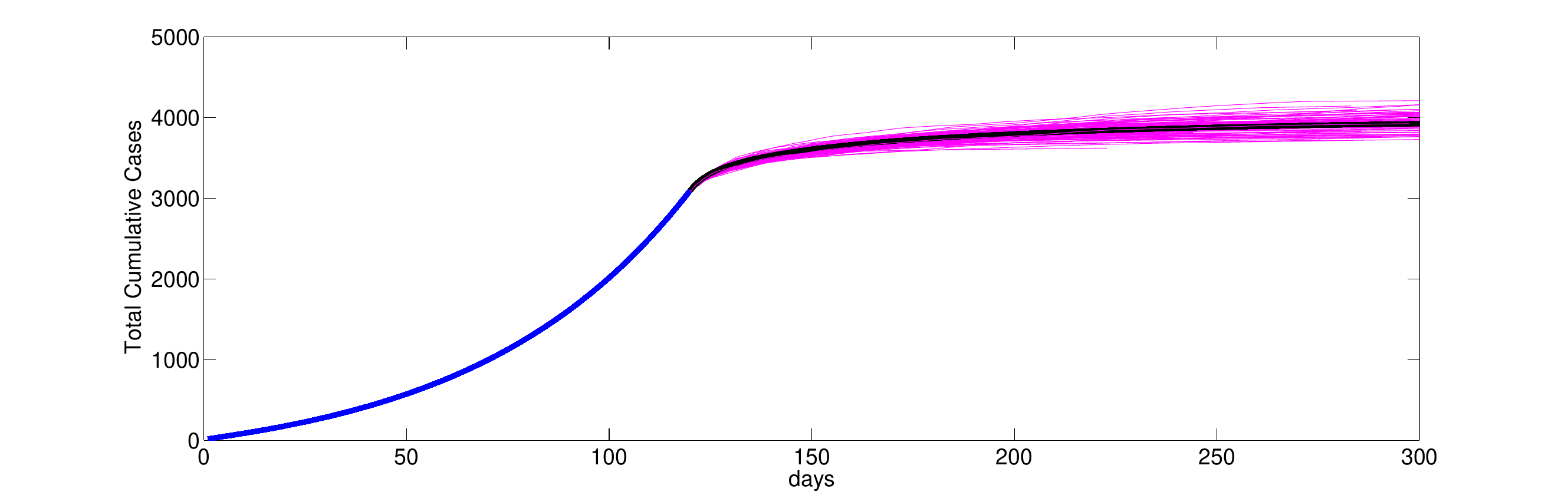}} 
 \caption{(a)  100 stochastic simulations of predicted cumulative reported cases in Sierra Leone forward from September 23, 2014, with $\alpha=0.2$, $\kappa=20$, $\pi_E=0.5$, $\pi_I=0.8$, $\omega_E=2000$, $\omega_I=1000$.  The stochastic simulations are in magenta and the ODE solution is in black.  (b) Simulations of predicted total cases corresponding to (a).  }
    \label{fig:mig2}
\end{figure}

\begin{figure}[!ht]
\subfigure[][]{\label{RepAlp12}\includegraphics[width=3.25in,height=2.0in]{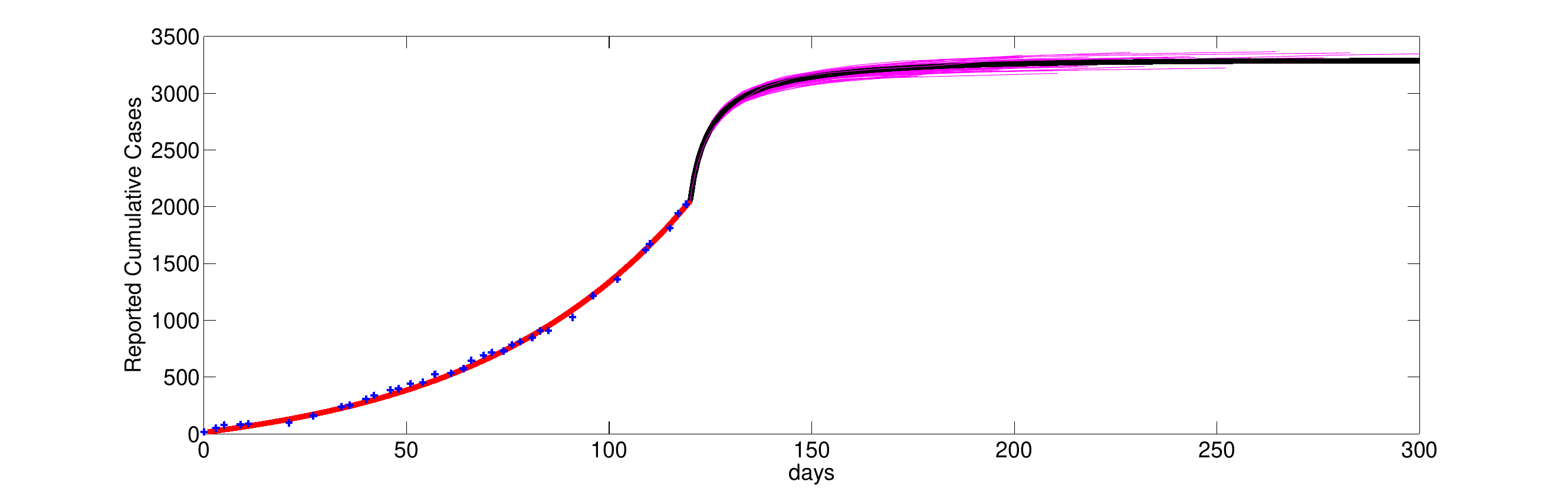}}
\subfigure[][]{\label{TotAlp12}\includegraphics[width=3.25in,height=2.0in]{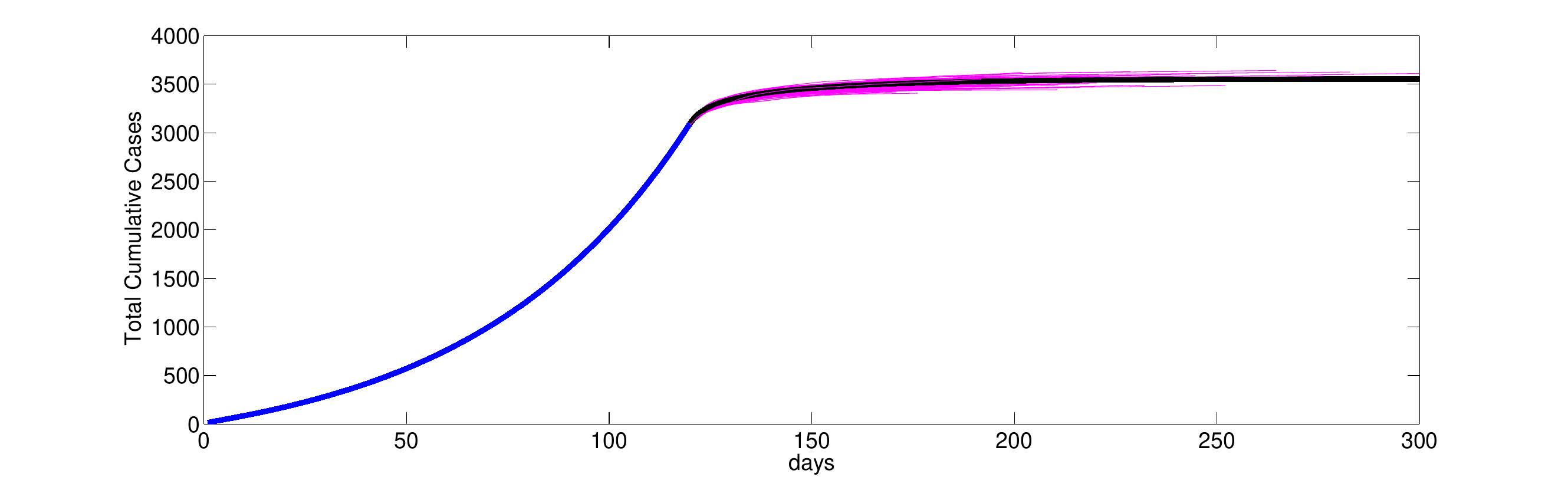}} 
  \caption{ (a)  100 stochastic simulations of predicted cumulative reported cases in Sierra Leone forward from September 23, 2014, with $\alpha=0.3$, $\kappa=20$, $\pi_E=0.5$, $\pi_I=0.8$, $\omega_E=2000$, $\omega_I=1000$.  The stochastic simulations are in magenta and the ODE solution is in black.  (b) Simulations of predicted total cases corresponding to (a). }
    \label{fig:mig3}
\end{figure}

\begin{figure}[!ht]
\subfigure[][]{\label{RepAlp101}\includegraphics[width=3.25in,height=2.0in]{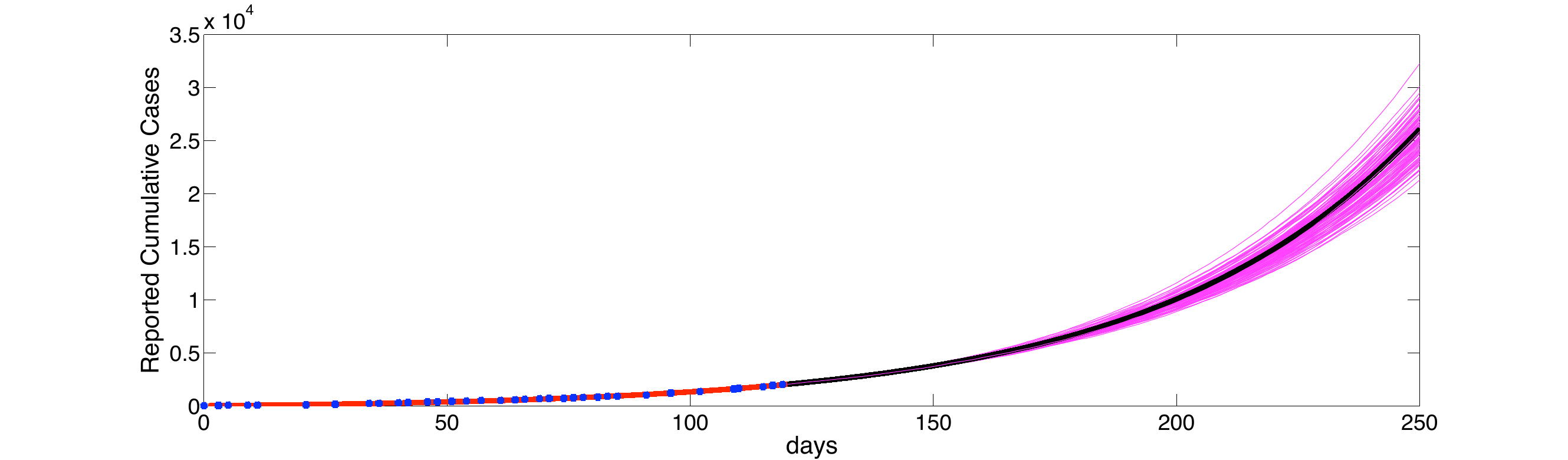}}
\subfigure[][]{\label{TotAlp101}\includegraphics[width=3.25in,height=2.0in]{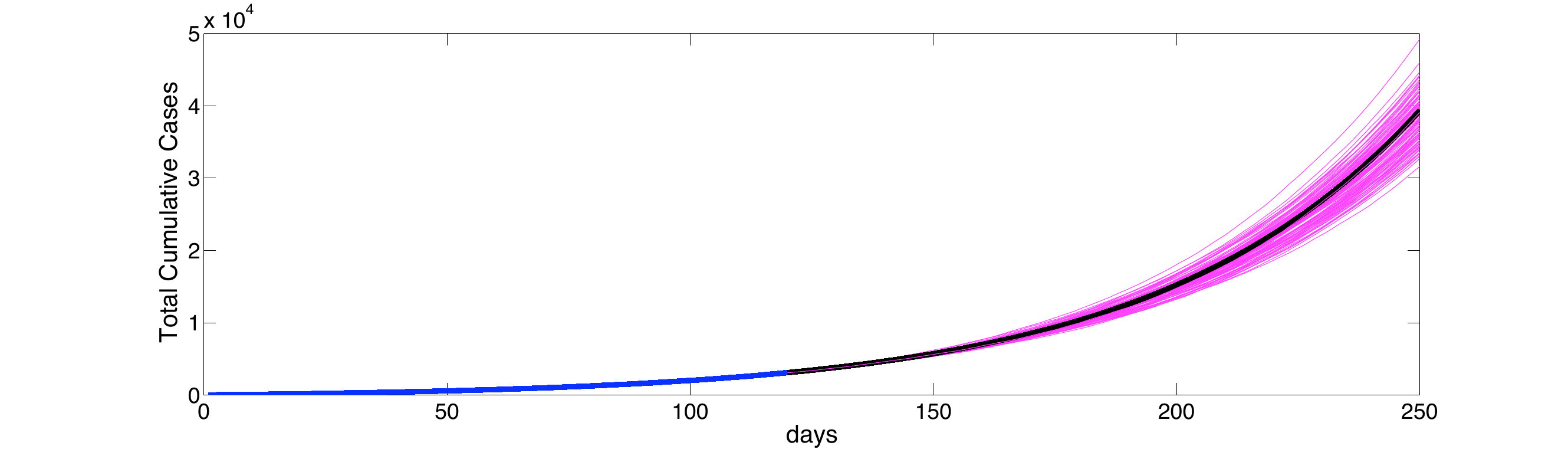}} 
  \caption{(a)  100 stochastic simulations of predicted cumulative reported cases in Sierra Leone forward from September 23, 2014, with $\alpha=0.1$, $\kappa=0$.  The stochastic simulations are in magenta and the ODE solution is in black.  (b) Simulations of predicted total cases corresponding to (a). }
    \label{fig:mig11}
\end{figure}

\begin{figure}[!ht]
\subfigure[][]{\label{RepAlp111}\includegraphics[width=3.25in,height=2.0in]{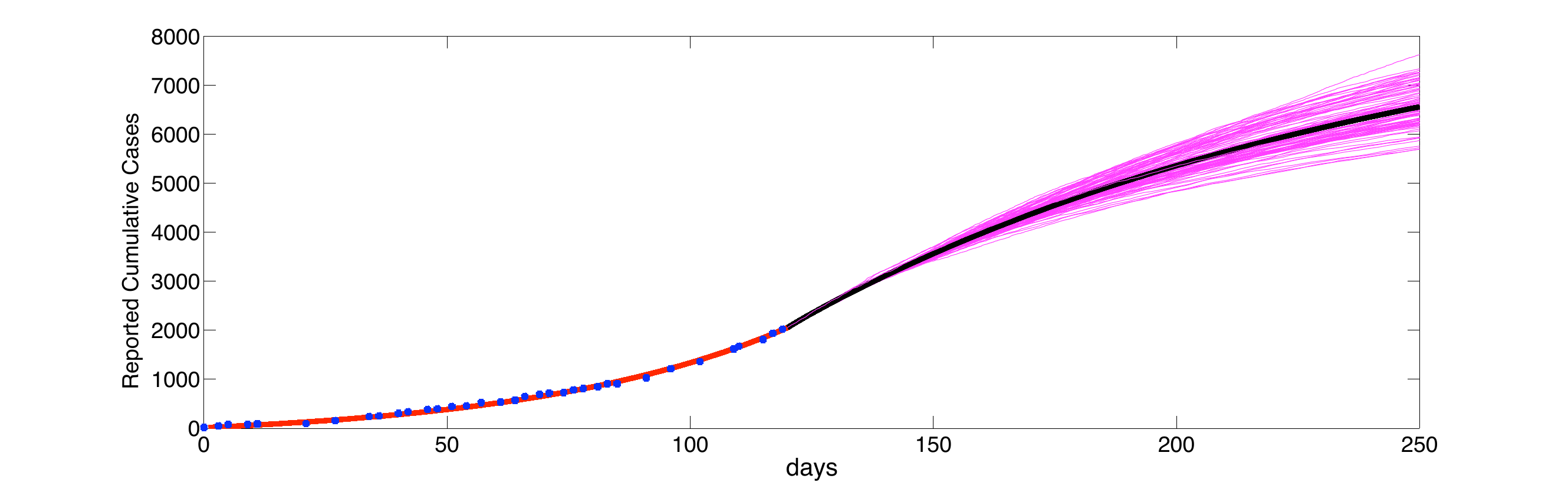}}
\subfigure[][]{\label{TotAlp111}\includegraphics[width=3.25in,height=2.0in]{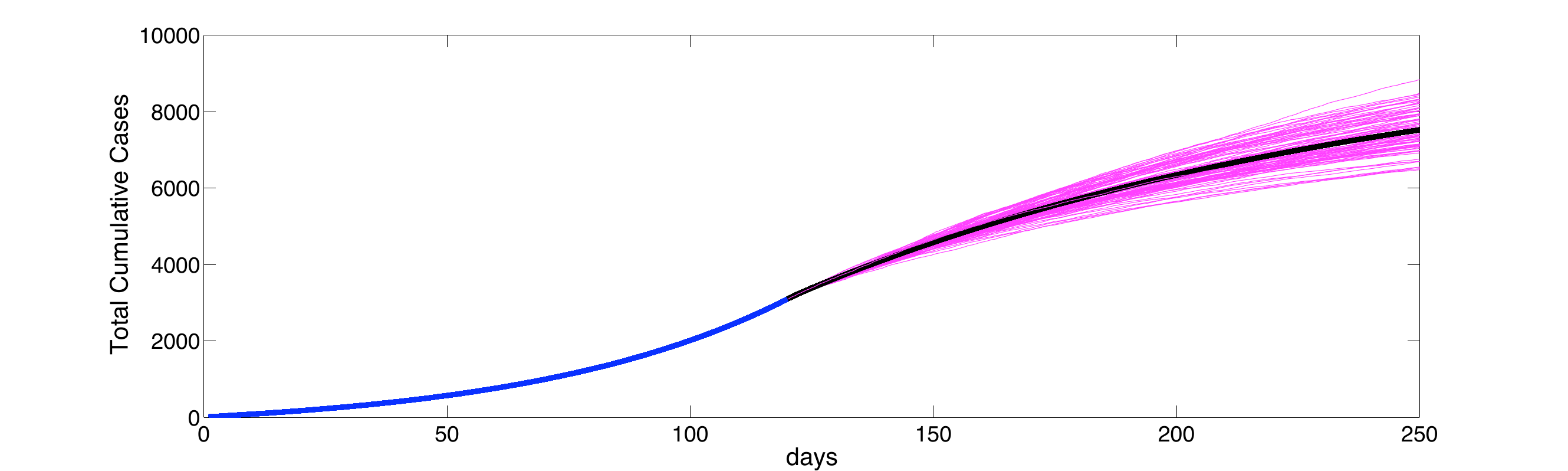}} 
 \caption{(a)  100 stochastic simulations of predicted cumulative reported cases in Sierra Leone forward from September 23, 2014, with $\alpha=0.2$, $\kappa=0$.  The stochastic simulations are in magenta and the ODE solution is in black.  (b) Simulations of predicted total cases corresponding to (a).  }
    \label{fig:mig21}
\end{figure}

\begin{figure}[!ht]
\subfigure[][]{\label{RepAlp121}\includegraphics[width=3.25in,height=2.0in]{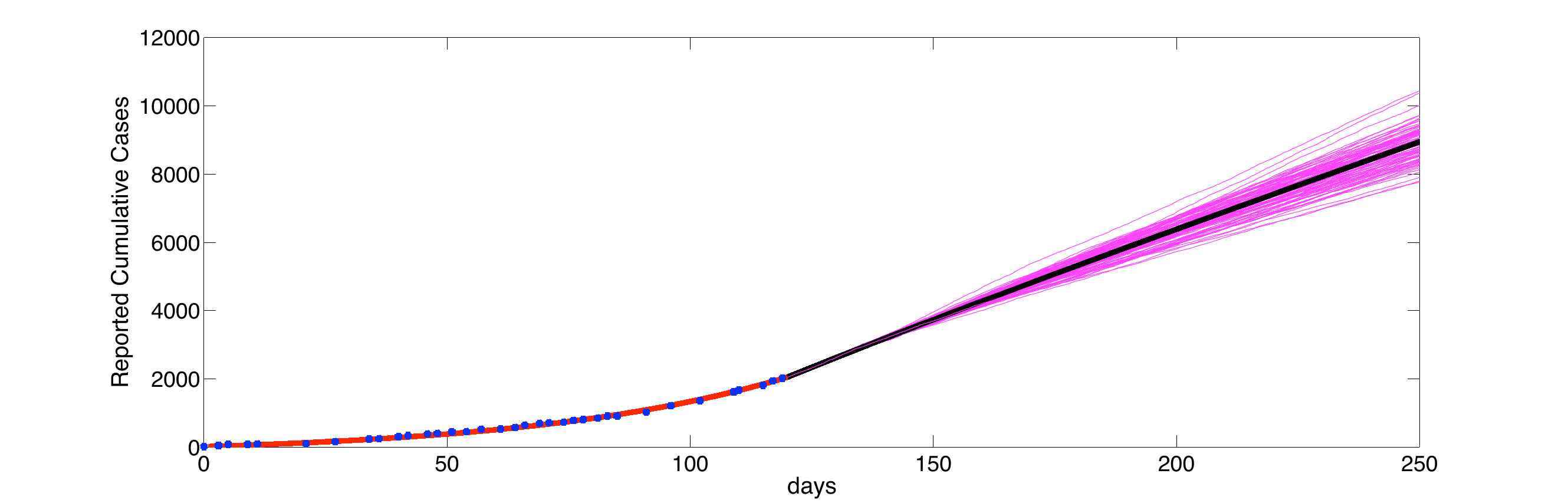}}
\subfigure[][]{\label{TotAlp121}\includegraphics[width=3.25in,height=2.0in]{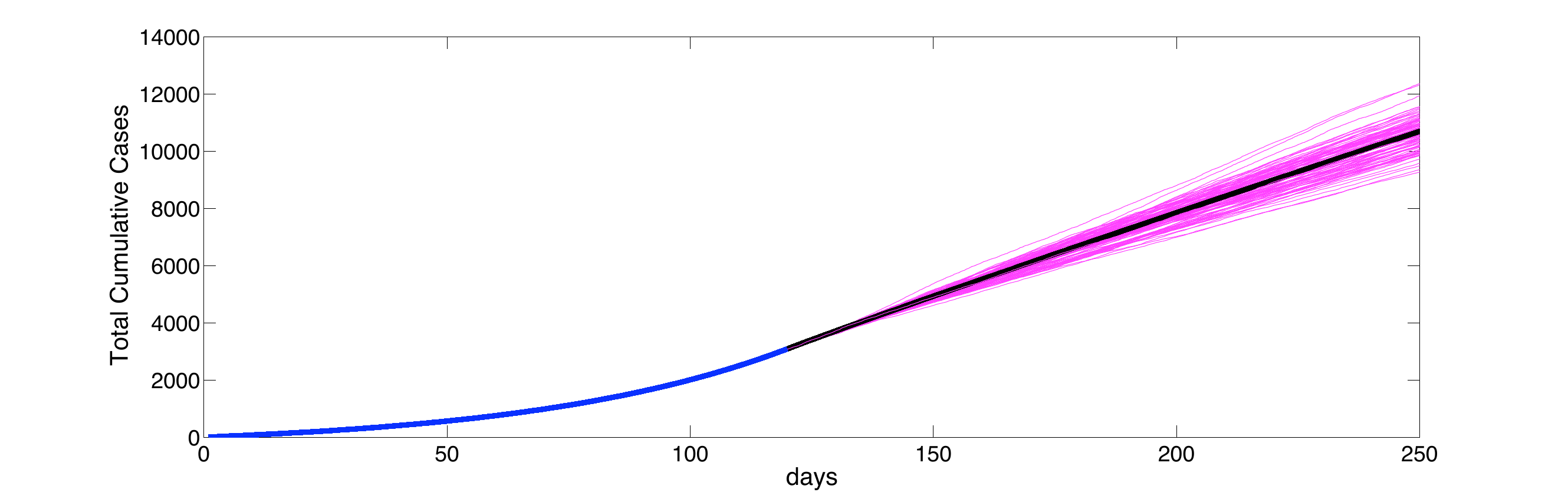}} 
  \caption{ (a)  100 stochastic simulations of predicted cumulative reported cases in Sierra Leone forward from September 23, 2014, with $\alpha=0.1$, $\kappa=10$, $\pi_E=0.1$, $\pi_I=0.5$, $\omega_E=2000$, $\omega_I=1000$.  The stochastic simulations are in magenta and the ODE solution is in black.  (b) Simulations of predicted total cases corresponding to (a). }
    \label{fig:mig31}
\end{figure}

\newpage
\begin{figure}[ht]
\begin{center}
 \includegraphics[width=6.5in,height=9.0in]{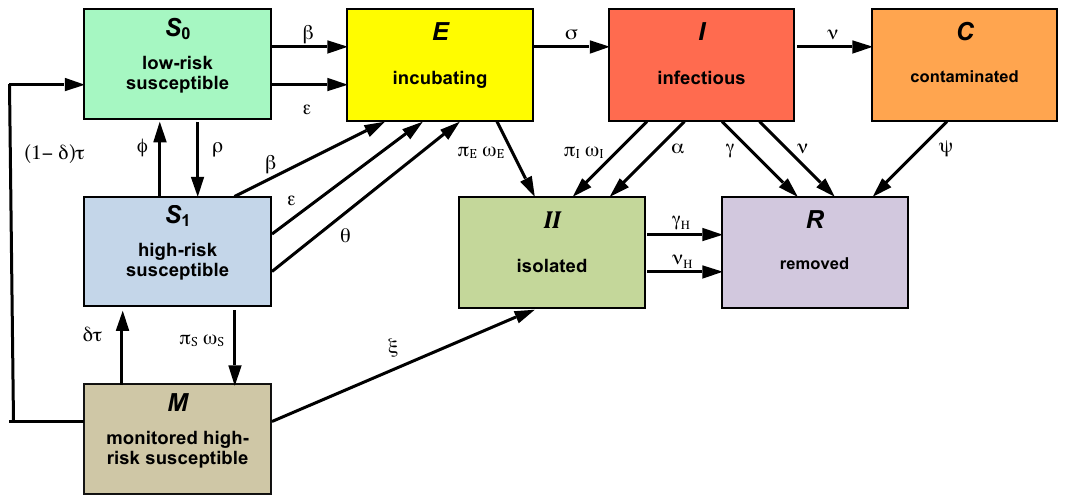}
\caption{Schematic diagram of the more general model compartments and  parameters.}\label{Fig2new}
\end{center}
\end{figure}

\end{document}